\documentclass[aps,pre,twocolumn]{revtex4-1}
\usepackage{amsmath}
\usepackage{epsfig}
\usepackage{graphicx}
\usepackage{array}
\usepackage{color}
\usepackage{amssymb}
\usepackage{epstopdf}
\usepackage{booktabs}
\usepackage[utf8]{inputenc}
\usepackage[T1]{fontenc}
\usepackage[colorlinks=True, linkcolor=blue, citecolor=blue]{hyperref}
\usepackage{float}
\usepackage{algorithm}
\usepackage{algpseudocode}

\begin{document}

\title{Hashing algorithms, optimized mappings and massive parallelization of multiconfigurational methods for bosons}
\author{Alex V. Andriati$^{1}$}\thanks{andriati@if.usp.br}
\author{Arnaldo Gammal$^1$}\thanks{gammal@if.usp.br}
\affiliation{$^{1}$Instituto de F\'{i}sica, Universidade de S\~ao Paulo, 05508-090 S\~ao Paulo, 
Brazil.}
\date{\today}

\begin{abstract}
    
Numerical routines for Fock states indexing and to handle creation and annihilation operators in the spanned multiconfigurational space are developed. From the combinatorial problem of fitting particles in a truncated basis of individual particle states, which defines the spanned multiconfigurational space, a hashing function is provided based on a metric to sort all possible configurations, which refers to sets of occupation numbers required in the definition of Fock states. Despite the hashing function unambiguously relates the configuration to the coefficient index of the many-particle state expansion in the Fock basis, averages of creation and annihilation operators can be a highly demanding computation, especially when they are embedded in a time-dependent problem. Therefore, improvements in the conversion between configurations after the action of creation and annihilation operators are thoroughly inspected, highlighting the advantages and additional memory consumption. We also exploit massive parallel processors from graphics processor units with CUDA to improve a routine to act with the many-body Hamiltonian matrix on the spanned multiconfigurational space, which demonstrated quantitatively the scalability of the problem. The improvements shown here seem promising especially for calculations involving a large number of particles, in which case, the optimized CUDA code provided a drastic performance gain of roughly fifty times faster than a single core processor. The codes were consistently tested with an application to the Lieb-Liniger gas, evaluating the ground state and comparing with the analytical solution.

\end{abstract}

\flushbottom

\maketitle


\section{Introduction}

\label{sec:intro}

Quantum many-particle problems can easily run out of analytical solutions when a few relevant assumptions are considered, such as the interactions among the particles. For bosonic systems, an example of analytically solvable model is the Lieb-Liniger(LL) gas~\cite{PhysRev.130.1605,PhysRev.130.1616}, consisting of a finite number of bosons confined in a periodic one-dimensional space. Despite the many-particle wave function can be obtained for the ground and excited states, the calculation of any observable is not trivial and require multi-dimensional integrals, with the dimension given by the number of particles.
    
After more than fifty years of the LL model has been reported, there is still ongoing research about it, for instance on approximations~\cite{SciPostPhys.3.1.003,PhysRevA.72.033613} and focusing on a narrow band of the excitation spectrum, to reproduce solitons predicted in the mean field theory~\cite{Sato_2016}. Besides, the hard core limit (impenetrable particles), which conducts to the Tonks-Girardeau(TG) gas~\cite{doi:10.1063/1.1703687,Yukalov_2005}, has been used in many one-dimensional studies as an upper bound when analyzing the effect of interaction strength~\cite{PhysRevA.72.033613,AnnaMinguzzi}.

The LL model illustrate well that in many particle physics there are many gaps in our understanding yet to be filled, from the fundamentals of quantum mechanics to collective phenomena, which cannot be thought by reductionism~\cite{Anderson1972}, even knowing the many-body wave function.

The importance of collective phenomena, concomitantly with the limitations of analytical approaches, justifies the progressive use of numerical computation, which has became almost indispensable on actual research. Surely, any numerical method employed impose limitations as well, though they generally are far less stiff than on a pure analytical approach.

Many particle systems are usually studied in terms of second quantization formalism in physics. An exact approach would require a complete set of Individual Particle States(IPS), also named orbitals\footnote{Methods employed here are also used in chemistry, from which came the designation for orbitals used on the study of molecules.}, which the particles can occupy. To overcome the drawback of an infinite dimensional space, a finite number of IPS are used instead, which turn possible to think on a numerical method. This IPS basis truncation and the spanned Fock basis are the core elements of multiconfigurational methods, where here, configuration means a possible arrangement of the particles in the IPS.

There are two approaches for the IPS, the first considering them fixed, like in the Bose-Hubbard model~\cite{AncientBoseHubbard,JAKSCH200552,PhysRevLett.81.3108,AncientSuperfluifInsulator,PhysRevB.58.R14741,PhysRevB.47.342}, and the second which establishes the IPS variationally minimizing the many-body action, including occasionally time dependence. In this case, we have the multiconfigurational time-dependent Hartree method, which started to be investigated firstly in molecular dynamics, in chemistry~\cite{10.1021/j100319a003,KOTLER1988483,MEYER199073,10.1063/1.459854,10.1063/1.463007}, and later in physics~\cite{MCTDHBderivation}, where many applications have been evaluated~\cite{HarmonicInteractModel,PhysRevA.94.063648,PhysRevA.92.033622,PhysRevA.97.043625,PhysRevLett.118.013603,PhysRevX.9.011052,PhysRevA.90.043620,PhysRevA.86.013630,PhysRevA.91.063621,PhysRevA.100.063625}. In any of the approaches for the IPS, a common requirement is to sort the configurations, so that the many-particle state can be expanded in the Fock basis with its coefficients properly enumerated. Moreover, in a numerical method, routines to convert between a configuration and its index are needed when using creation and annihilation operators.

In this work, we establish a one-to-one mapping between configurations and integer numbers, indexes of coefficients of many-particle state expansion in the Fock basis, studying the performance of the routines for bosons. Improvements through direct mappings between configurations whose the occupation numbers differ by the action of creation and annihilation operators are proposed, which are very relevant to methods that must evaluate average of these operators or compute the action of many-particle Hamiltonian matrix several times. For instance, Exact Diagonalization(ED)~\cite{Weibe2008,Zhang_2010,PhysRevB.42.6561,Ravent_s_2017,Sandvik} and the Multiconfigurational Time-Dependent Hartree method for Bosons(MCTDHB)~\cite{MCTDHBderivation} require computation of these quantities several times either to obtain stationary states or study dynamics. Finally the scalability of the Hamiltonian matrix action in the multiconfigurational space is studied using multi-core CPU and GPU, which yields drastic improvements for a large number of particles.

\section{The multiconfigurational space and relevant operators}

Quantum and statistical mechanics often resort to second quantized formalism, specially when dealing with a system of identical particles, which automatically takes into account the symmetry of the many-particle wave function. In this formalism, all observables can be expressed in terms of creation and annihilation operators, defined by two possible algebras
\begin{equation}
[ \hat{a}_k, \hat{a}^{\dagger}_l ] = \delta_{kl} ,
\end{equation}

\noindent for bosons and
\begin{equation}
\{ \hat{c}_k, \hat{c}^{\dagger}_l \} = \delta_{kl} ,
\end{equation}

\noindent for fermions, where $[A,B] = AB - BC$ and $\{A,B\} = AB + BA$. Our focus here are bosons as outlined in the introduction.

In the formalism is required a complete set of IPS, whose the creation/annihilation index refers to. As mentioned before, instead of a complete set, a generic finite set of IPS $\{ | \phi_k \rangle \}_{k \in \{1,...,M\}}$ is needed for multiconfigurational methods, independent whether they are obtained from a variational approach or not. With these definitions, a non-interacting many-particle operator is written as
\begin{equation}
\hat{\mathcal{T}} = \sum_{k,l} \hat{a}^{\dagger}_k \hat{a}_l \langle \phi_k | T | \phi_l \rangle ,
\label{eq:noninteracting}
\end{equation}

\noindent and the interacting many-particle operators as
\begin{equation}
\hat{\mathcal{V}} = \frac{1}{2}\sum_{k,l,q,s} \hat{a}^{\dagger}_k \hat{a}^{\dagger}_s
\hat{a}_l \hat{a}_q
\langle \phi_k , \phi_s | V | \phi_q , \phi_l \rangle .
\label{eq:interacting}
\end{equation}

\noindent Moreover, Hamiltonians are written as a combination of both, with $\hat{\mathcal{H}} = \hat{\mathcal{T}} + \hat{\mathcal{V}}$. Here, and throughout this text, the many-body operators in second quantized form are denoted with a hat, while uppercase letters without hat are used for ordinary one and two-body operators. Besides, we use lowercase Greek letters for the IPS and uppercase Greek letters are reserved for many-particle states.

The many-particle state can in principle be written as a linear combination of Fock states, which are simultaneous eigenstates of $\hat{a}^{\dagger}_k \hat{a}_k$ for every $k \in \{1, ...,M\}$ that label the IPS, with the corresponding eigenvalue being the number of particles occupying the IPS $k$. A Fock state is denoted throughout the text by
\begin{equation}
| \vec{n} \rangle \doteq | n_1 \ ... \ n_M \rangle \ , \ \mathrm{such \ that} \quad \hat{a}^{\dagger}_k \hat{a}_k | \vec{n} \rangle = n_k | \vec{n} \rangle ,
\label{eq:FockState}
\end{equation}

\noindent where $\vec{n}$ must be a valid configuration, what means that the sum of occupations must give the total number of particles. Thus any configuration $\vec{n}^{(\beta)}$ must satisfy
\begin{equation}
\sum_{k=1}^M n_k^{(\beta)} = N ,
\label{eq:constraint}
\end{equation}

\noindent where $N$ is the total number of particles.

In bosonic systems, the occupations in any state $k$ goes from $0$ to $N$ respecting the condition~\eqref{eq:constraint}. Therefore, the total number of configurations $\vec{n}^{(\beta)}$ is obtained from the combinatorial problem of how to fit $N$ identical balls in $M$ boxes, which yields for bosons
\begin{equation}
N_c(N,M) = \binom{N + M - 1}{M - 1} = \frac{(N + M - 1)!}{N! ( M - 1 )!} .
\label{eq:nc}
\end{equation}

\noindent Consequently, the configurations can be indexed by $\beta \in \{1,2,...,N_c(N,M)\}$. Since there are many ways to select the occupation numbers under the constraint of Eq.~\eqref{eq:constraint}, expressing the many-body state in this basis is called a multiconfigurational method. Therefore, introducing $\mathbb{H}(N,M)$ for the multiconfiguration space of $N$ particles and $M$ IPS yields
\begin{multline}
\mathbb{H}(N,M) = \mathrm{span} \Bigg\{| \vec{n}^{(\beta)} \rangle \ ; \ \sum_{j=1}^M n_j^{(\beta)} = N \ , \\ 
\forall \beta \in \{ 1,...,N_c(N,M) \} \Bigg\} .
\end{multline}

In this multiconfigurational space, we can express the many-particle state of the system by a linear 
combination as
\begin{equation}
| \Psi(t) \rangle = \sum_{\beta=1}^{N_c(N,M)} C_{\beta}(t) | \vec{n}^{(\beta)} \rangle ,
\end{equation}

\noindent where $\vec{C}(t)$ is a complex vector of dimension $N_c(N,M)$.

Independently of how the IPS are selected, average of the operators in Eq.~\eqref{eq:noninteracting} and \eqref{eq:interacting} need to be evaluated in the multiconfigurational basis, which require some way to act with creation/annihilation operators on the Fock states $| \vec{n}^{(\beta)} \rangle$ numerically. This problem is studied in this paper, starting from the fundamental question on how to establish the relation between $\beta$ and its configuration $\vec{n}^{(\beta)}$. A function that does this job is called a hashing function and we have a perfect hashing if we get a one-to-one correspondence. In the last three decades at least, some effort was directed in developing hashing functions~\cite{PhysRevB.42.6561,LIANG199511}, although most part of more recent works have been directed to particles confined in sites of optical lattices~\cite{Ravent_s_2017}, restricting to the Bose-Hubbard model~\cite{Zhang_2010} which restrict the interactions to particles in the same site, or for systems with just spin as degree of freedom~\cite{PhysRevB.42.6561,Sandvik,JIA201881}. A general study about the time demanded when computing the averages mentioned above and improvements was not performed so far.

The physical operators chosen to illustrate the performance throughout this paper are the one- and two-body density matrices and the many-body Hamiltonian. These quantities are essential for the MCTDHB~\cite{MCTDHBderivation}, where usually have to be evaluated several times for the same configurational space. Other applications are for the low-lying excited states or the ground state with ED~\cite{Weibe2008}, where the use of iterative methods as Lanczos tridiagonal decomposition~\cite{Lanczos:1950,Loan,Demmel.ch7} requires many Hamiltonian matrix multiplications. Moreover, in time-dependent problems, the use of Short Iterative Lanczos(SIL) integrator is recurrent~\cite{MeyerSpringer1997,ParkLight}.

The one- and two-body matrices are given by the expectation values of a combination of two or four creation and annihilation operators as
\begin{equation}
\rho^{(1)}_{kl} = \langle \Psi | \hat{a}^{\dagger}_k \hat{a}_l | \Psi \rangle = \!
\sum_{\substack{\gamma=1,\\\beta=1}}^{N_c(N,M)} \!\! C_{\gamma}^{*} C_{\beta} \langle 
\vec{n}^{(\gamma)} | \hat{a}^{\dagger}_k \hat{a}_l | \vec{n}^{(\beta)} \rangle ,    
\label{eq:rho1}
\end{equation}

\noindent and
\begin{multline}
\rho^{(2)}_{klqs} = \langle \Psi | \hat{a}^{\dagger}_k \hat{a}^{\dagger}_l \hat{a}_q
\hat{a}_s | \Psi \rangle = \\
\sum_{\substack{\gamma=1,\\\beta=1}}^{N_c(N,M)} C_{\gamma}^{*} C_{\beta} \langle 
\vec{n}^{(\gamma)} |
\hat{a}^{\dagger}_k \hat{a}^{\dagger}_l \hat{a}_q \hat{a}_s | \vec{n}^{(\beta)} \rangle ,
\label{eq:rho2}
\end{multline}

\noindent respectively. In appendix~\ref{appendix-matrixelements}, the different rules to compute the elements of $\rho^{(1)}$ and $\rho^{(2)}$ are summarized.

The Hamiltonian is an important operator to study dynamics or the ground state in different methods as mentioned before. Therefore, the time demanded to apply it using the multiconfigurational space is evaluated here by measuring the time elapsed to set up the output vector $\tilde{\mathbf{C}}$ from
\begin{equation}
\tilde{\mathbf{C}} = \mathcal{H} \cdot \mathbf{C} \ , \quad \tilde{C}_{\gamma} = \sum_{\beta=1}^{N_c(N,M)} \!\! \mathcal{H}_{\gamma \beta} C_{\beta} \ ,
\label{eq:Haction}
\end{equation}

\noindent where $\mathcal{H}$ is the matrix representation of $\hat{\mathcal{H}}$ in the multiconfigurational space with $\mathcal{H}_{\gamma \beta} = \langle \vec{n}^{\gamma} | \hat{\mathcal{H}} | \vec{n}^{\beta} \rangle$, and $\hat{\mathcal{H}}$ is a combination of Eqs.~(\ref{eq:noninteracting},\ref{eq:interacting}),

\begin{multline}
\hat{\mathcal{H}} = \sum_{l, k = 1}^{M} \Bigg[
\langle \phi_l | T | \phi_k \rangle \hat{a}^{\dagger}_l \hat{a}_k
\\ + \frac{1}{2} \sum_{s, q = 1}^{M}
\hat{a}_l^{\dagger} \hat{a}_k^{\dagger} \hat{a}_s \hat{a}_q
\langle \phi_l , \phi_k | V | \phi_q, \phi_s \rangle
\Bigg] \ .
\end{multline}

The Hamiltonian matrix $\mathcal{H}$ is very sparse, however, it is not stored in our implementations. The clear advantage in not storing the Hamiltonian matrix appears when dealing with a method that needs to update the IPS  $\{| \phi_k \rangle \}$, as in the MCTDHB because they are time-dependent, which would require to reassemble the matrix during the time evolution. There is also interest when the goal is to diagonalize using the Lanczos algorithm, where one only needs a routine to apply the matrix on a vector and not the matrix itself.

We provide a supplemental material with codes in C language used to collect data in this paper, with a program summary described in appendix~\ref{app:programs}. The files that include routines to compute $\rho^{(1)}$, $\rho^{(2)}$ and to apply the Hamiltonian on the coefficients are provided in ``onebodyMatrix.h'', ``twobodyMatrix'' and ``hamiltonianMatrix.h'' respectively.

\section{Mapping Fock states to integers}
\label{sec:sec2}

Following the theoretical framework developed in the previous section, as a first step, it is 
necessary to address an integer number for each configuration. The routine to perform this task 
assigns a cost for every IPS to be occupied, starting with all occupations zero. The problem can be 
depicted by a basket of balls (particles), that starts with $N$, the total number of particles, and is emptied to fill the IPS. The enumerations for now on will start from $0$, which is more convenient in the numerical approach.

Given an enumeration for the IPS from $0$ to $M - 1$, if we take  one with number $k$ where $0 
\leq k \leq M - 1$, there are other $k$ IPS below it since we are counting the number zero. In 
this way, the cost to put one particle in the state $k$ is defined by the total number of
configurations of remaining particles in the basket over all previous IPS. In other words, the cost 
is all the combinations we could do with the lower number IPS and particles in the basket. When the 
basket has none particle left, the process is finished and the total cost will be the index of the 
configuration.

\begin{figure}[t]
    \centering
    \includegraphics[scale=0.75]{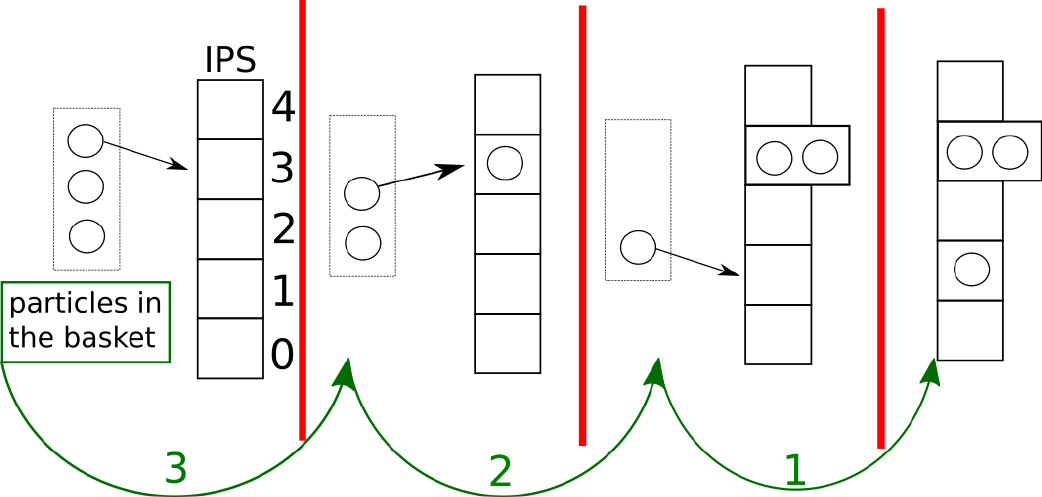}
    \caption{Example to illustrate the process of IPS occupation for $N = 3$ and $M = 5$ for a 
        the specific configuration $|0,1,0,2,0\rangle$. The total cost of this configuration is 
        $N_c(3,3) + N_c(2,3) + N_c(1,1) = 17$ where the terms are presented following the order of 
        arrows.}
    \label{fig:scheme2}
\end{figure}

In Fig.~\ref{fig:scheme2} is depicted a practical example of the description above. The combination 
function defined in Eq.~\eqref{eq:nc} plays a crucial role being used to compute the costs. For 
instance, if $p_n$ is the IPS the $n$-th particle occupies, where by construction we have $0 \leq 
p_1 \leq ... \leq p_N < M$, then the total cost mentioned above can be compute by

\begin{multline}
I(\vec{p}) = \sum_{n = 1}^{N} N_c(n,p_n) = \sum_{n = 1}^{N} \binom{n + p_n - 1}{p_n - 1} , \\ \mathrm{with} \ N_c(n,0) = 0, \ \forall n .
\label{eq:TotalCost}
\end{multline}

The Eq.~\eqref{eq:TotalCost} is identical to the results in Ref.~\cite{LIANG199511}, though some conventions are changed and there the derivation follows an alternative way, using a correspondence to fermions. Moreover from Ref.~\cite{LIANG199511}, it is already known that this relation maps uniquely integers to configurations without any left number and therefore indicates a perfect hashing function.

\begin{figure}[b]

\begin{algorithm}[H]
    
    \caption{Get Index from a configuration}
    
    \begin{algorithmic}
        
        \Require $\ N > 0 , \ M > 0 , \ \mathrm{occupation\ vector}\ \vec{n}$
        
        \State $k \leftarrow 0$
        \State $s \leftarrow N$
        
        \For{$m = M-1 .. 1$}
        \State $j \leftarrow n[m]$
        
        \While{$j > 0$}
        \State $k \leftarrow k + N_c(s,m)$
        \State $s \leftarrow s - 1$
        \State $j \leftarrow j - 1$
        \EndWhile
        \EndFor
        
        \noindent \Return $k$
        
    \end{algorithmic}
    
    \label{alg:conf2ind}
    
\end{algorithm}

\end{figure}

\begin{figure}
    
\begin{algorithm}[H]
    
    \caption{Build configuration $\vec{n}$ from index $\beta$}
    
    \begin{algorithmic}
        
        \Require $\ N > 0 , \ M > 0 , \ N_c(N,M) > \beta \geq 0$
        
        \For{$i = 0 .. M-1$}
        \State $n[i] \leftarrow 0$
        \EndFor
        
        \State $k \leftarrow \beta$
        \State $m \leftarrow M - 1$
        \State $s \leftarrow N$
        
        \While{$k > 0$}
        
        \While{$k - N_c(s,m) < 0$}
        \State $m \leftarrow m - 1$
        \EndWhile
        
        \State $k \leftarrow k - N_c(s,m)$
        \State $n[m] \leftarrow n[m] + 1$
        \State $s \leftarrow s - 1$
        
        \EndWhile
        
        \If{$s > 0$} \State $n[0] \leftarrow n[0] + s$ \EndIf
        
    \end{algorithmic}
    
    \label{alg:ind2conf}
    
\end{algorithm}

\end{figure}

The prescription of the hashing can now be implemented. Given a configuration, to discover its 
index, we sum up the costs, removing particle by particle using Eq.~\eqref{eq:TotalCost}. This procedure is detailed in Algorithm~\ref{alg:conf2ind}. The reverse process is quite straightforward, to assemble the configuration given an index between $0$ and $N_c(N,M) - 1$, we need to put all the 
particles in a basket and check, starting from the IPS $M - 1$, if the index is bigger than the 
cost to put a particle. In positive case, a particle is transferred from the basket to the IPS,  otherwise move to lower cost IPS and try again. The routine to convert index to configuration is  given in Algorithm \ref{alg:ind2conf}. The algorithms \ref{alg:conf2ind} and \ref{alg:ind2conf} are the core functions to operate with creation and annihilation given a many-body state in the multiconfigurational basis.

\begin{figure*}[t]
    \centering
    \includegraphics[scale=1.0]{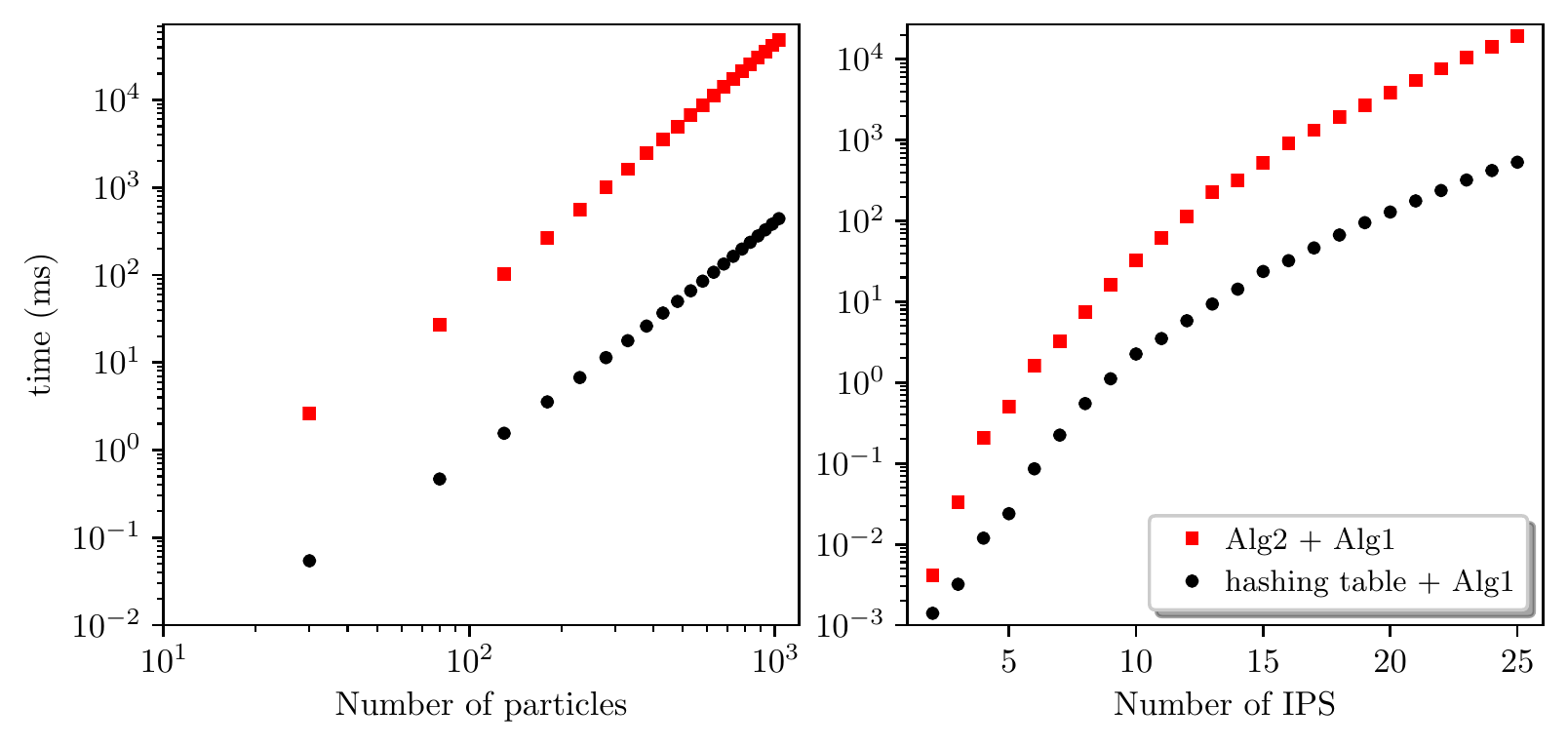}
    \caption{Time  to compute all elements of $\rho^{(1)}$. The red 
        squares correspond to an implementation that uses just the conversion algorithms~\ref{alg:ind2conf} and \ref{alg:conf2ind} and the black circles make use of a hashing table to store and sort the configurations, which restrict to use only algorithm \ref{alg:conf2ind}. In the left panel was fixed three IPS while varying the number of particles and in the right panel was varied the number of IPS with five particles.}
    \label{fig:trho1}
\end{figure*}

In the computation of the density matrices, from Eqs.~\eqref{eq:rho1} and \eqref{eq:rho2}, we need to perform just the sum in $\beta$, whereas for each $\beta$ there is a unique value for $\gamma$, the one corresponding to the configuration after replacing the particles due to the action of the creation and annihilation operators. Therefore, for $\beta$ running from $0$ to $N_c(N,M) - 1$, we need three steps to perform the operation required. First, obtain the configuration using algorithm \ref{alg:ind2conf}. Second, reconfigure the occupation according to the action of creation/annihilation operators. Third, use this new occupation vector to compute the corresponding index $\gamma$ using algorithm \ref{alg:conf2ind} and do the multiplication of coefficients with the additional rules listed in appendix~\ref{appendix-matrixelements}.

Nevertheless, we must use the algorithms \ref{alg:ind2conf} and \ref{alg:conf2ind} $N_c(N,M)$ times 
for every element of the density matrices, that results in a total of $M^2 N_c(N,M)$ calls of both 
functions to setup all the elements of $\rho^{(1)}$ and $M^4 N_c(N,M)$ for 
$\rho^{(2)}$~\footnote{Actually, this number can be halved if one uses hermiticity, and reduced even more using the commutation relations in the case of $\rho^{(2)}$. Moreover, when the indexes of the creation and annihilation operators are the same no calls of the conversion algorithms are need at all, since there is no rearrangement of particles.}. However, we may spend more memory creating some structures to avoid the number of calls of these functions, which will improve performance as will be shown later. Surely, the setup of any new structure would demand some equivalent time but, we again emphasize that our goal are problems that need to compute these quantities several times for the same configurational space \cite{MCTDHBderivation}.

A first improvement is to build once all occupation vectors and maintain them stored during all 
operations, defining a hashing table. For instance, they can be stored along rows of a matrix of 
integers, with the row number being the index of the respective configuration. This hashing table 
would require to store $M N_c(N,M)$ integers in exchange of avoiding calls of algorithm 
\ref{alg:ind2conf} when computing $\rho^{(1)}$ and $\rho^{(2)}$.

The time required to set all $\rho^{(1)}$ elements is shown in Fig.~\ref{fig:trho1} using two different implementations, the first using both algorithms \ref{alg:ind2conf} and \ref{alg:conf2ind}, and the second that uses hashing table(all configurations previously defined) and algorithm \ref{alg:conf2ind}. The basic difference between the two implementations is the call of algorithm \ref{alg:ind2conf} face to a memory access of the hashing table. As can be noted in Fig.~\ref{fig:trho1}, the performance gain is critical, highlighted by the logarithm scale, for both cases, when varying the number of particles or the number of IPS. Moreover for the left panel, the time required with respect to the number of particles has a clear linear relation in logarithmic scales, which indicates a power law of the form $\tau = b N^a$, with $\tau$ the time and $N$ the number of particle, which will be investigated later

The density matrix $\rho^{(1)}$ is computed in Fig.~\ref{fig:trho1} by randomly generating the components of the vector $\vec{C}$, normalizing it to one and using Eq.~\eqref{eq:rho1} with the rules detailed in~\ref{appendix-matrixelements}. Codes in C language are available in the supplemental material. In the ``configurationsMap.h'' file, the function named ``setupFocks'' set the hashing table of configurations, while the functions ``FockToIndex'' and ``IndexToFock'' correspond to algorithms \ref{alg:conf2ind} and \ref{alg:ind2conf} respectively. Besides, both routines to assemble $\rho^{(1)}$ are provided in ``onebodyMatrix.h'' file. Further description about the codes and the program to measure the time elapsed is given in appendix~\ref{app:programs}

Other routines, to build the two-body density matrices and to apply the Hamiltonian using the 
configuration basis shall benefit even more from the hashing table, because they require much more 
operations. We thus focus on these two quantities for the next improvements.

\section{Mapping routines}

The next step is to set the routines to compute the operators of Eqs.~\eqref{eq:rho1}, \eqref{eq:rho2} and  \eqref{eq:Haction} completely free from calls of the algorithm \ref{alg:conf2ind} as well. For this aim, it is necessary to define a structure where given an index it has stored all possible jumps~\footnote{Jump here means the simultaneous destruction and creation of particle in different states.} of one and two particles among the IPS, corresponding to the action of one and two pairs of creation/annihilation operators respectively.

In a single particle jump, one has (at most)$M$ different states to remove a particle and $M$ 
different states to place it back, which implies that for every configuration there are at most 
$M^2$ possible transitions. Thus a straightforward way to map all these transitions is to define a 
triple indexed structure, which stores integers, where the first index is from the configuration 
number, and the other two are IPS numbers, one from where the particle is being destroyed and other 
where it is being created. This one-particle jump mappings would require $N_c(N,M) M^2$ new integers to store. In the supplemental material the function ``OneOneMap'' in ``configurationsMap.h'' file implements in C code this mapping using an array of integers.

It is worth pointing out that the memory cost for this one-particle jump mapping is greater than the 
first improvement of the hashing table, where in that case was stored all the occupation numbers 
and therefore had a cost of $N_c(N,M) M$ integers. This justify why we did not mind about memory 
cost at that stage. Moreover, the $N_c(N,M) M^2$ integers wastes some memory because there are configurations with some empty IPS, which actually do not have $M^2$ possible transitions. Nevertheless, this wasted memory here will not matter, since the two-particle jump mappings will require more elements than $N_c(N,M) M^2$, as will be shown later.

We now analyze how to implement a structure that maps double jumps, that is, two particles move to 
different IPS. If we follow the same idea presented for the one-particle jump, for 
each configuration there would be at most $M^2$ possibilities to take two particles from the 
occupation numbers and for each one of these possibilities there are again $M^2$ ways to replace 
them. Following this naive way, we would end up with the additional memory requirement of $N_c(N,M) 
M^4$ integers. Nevertheless, it is possible to reduce this number.

A thoroughly inspection over configurations show us that $M^2$ possibilities to remove two particles from the IPS (equivalent to the action of two annihilation operators) is not true for some configurations, because there are many configurations with empty IPS. The real number of possibilities can be obtained as follows: for every non-empty IPS $k$ we search for $s \geq k$ non-empty as well, and whenever we find such numbers, we will have $M^2$ possible IPS to replace these particle taken from $k$ and $s$.

\begin{figure}[b]
    \centering
    \includegraphics[scale=0.55]{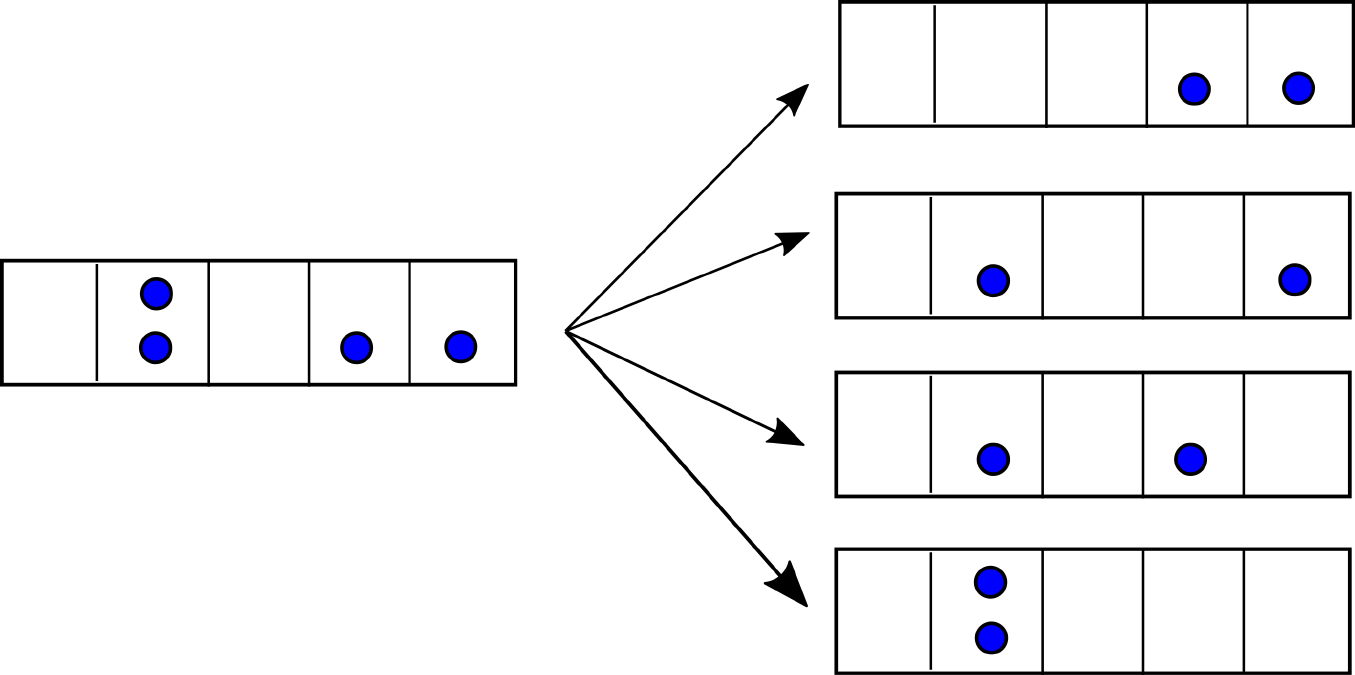}
    \caption{In the left a possible configuration for $M = 5$ and $N = 4$. In the right the arrows indicate all possible ways to remove 2 particles from the single particle states, that is 4.}
    \label{fig:jump_illustration}
\end{figure}

In Fig.~\ref{fig:jump_illustration}, it is illustrated for a simple case, given a specific 
configuration, the possible ways to remove simultaneously two particles by the action of two 
annihilation operators. The arrows conducts to the possible outcomes, where for each one, we have 
$M^2$ possibilities to replace the particles. Originally, the naive way would store a lot of useless information since it considers a bunch of forbidden transitions, that is, removal from empty states. For instance, in the case represented in Fig.~\ref{fig:jump_illustration}, it would require $5^4 = 625$ possibilities, while there are only $4 \times 5^2 = 100$ real possibilities.

In summary, to save memory for this structure of double jump mappings, we cannot allocate those forbidden transitions. In order to overcome the problem, we define a structure like a hashing table, though each line of the table has a variable number of elements, where the line number correspond to an index of a configuration. Its elements are integers, indexes of other configurations that are outcomes of all possible jumps of two particles.

A possible way to sort the elements for each line in the table is starting with the IPS $k = 0$ up to $k = M - 1$, we take $k \leq s < M$ and for each possible simultaneous removing of particles in $k$ and $s$, we have a stride of $M^2$ integer numbers that corresponds to new configurations obtained for every possible way to replace the particles removed. Therefore, if one wants to know the configuration index $\gamma$, that is a result of rearranging two particles in another configuration $\vec{n}^{(\beta)}$, removing from IPS $i$ and $j \geq i$ and replacing them in $q$ and $l$ IPS, it is required to check out how many strides must be ignored. In this case, the number of strides is the number of possible simultaneous removal from IPS $k$ and $s$ for every $k = 0, ..., i$ and $s = k, ... , j - 1$.

For example, suppose in Fig.~\ref{fig:jump_illustration} we are interested in the transitions that 
come from removing the last two particles, thus we need to skip 3 strides. In other words, in our 
table, in the line corresponding to the configuration in the figure, we need to skip the $3 \times 
5^2$ elements to get the indexes of configurations we are interested in this particular example.

The implementation for the two-jump mapping described above is done by splitting the problem in two parts, as can be consulted in the supplemental material. The first mappings refers to the annihilation of two particles from the same IPS, and is implemented by the function ``OneTwoMap'' in ``configurationsMap.h'' file. The second refers to the annihilation of 2 particle in necessarily different IPS and is implemented in ``TwoTwoMap'' function in the same file. This procedure avoids conditional statements when computing the number of strides, and is suitable for the different rules presented in the appendix~\ref{appendix-matrixelements}.

In the following we compare the performance between implementations that uses only hashing table and the ones that uses jump mapping. Before moving on, in the  C code files in the supplemental material, the ``demonstrateFockMap.c'' can be executed to see the basic functionality of the structures described so far. Its execution is rather simple and requires only the number of particles and number of IPS as command line arguments to print the hashing table and some random jumps, which are addressed to other configuration using the mappings described.

\begin{figure*}[t]
    \centering
    \includegraphics[scale=1.0]{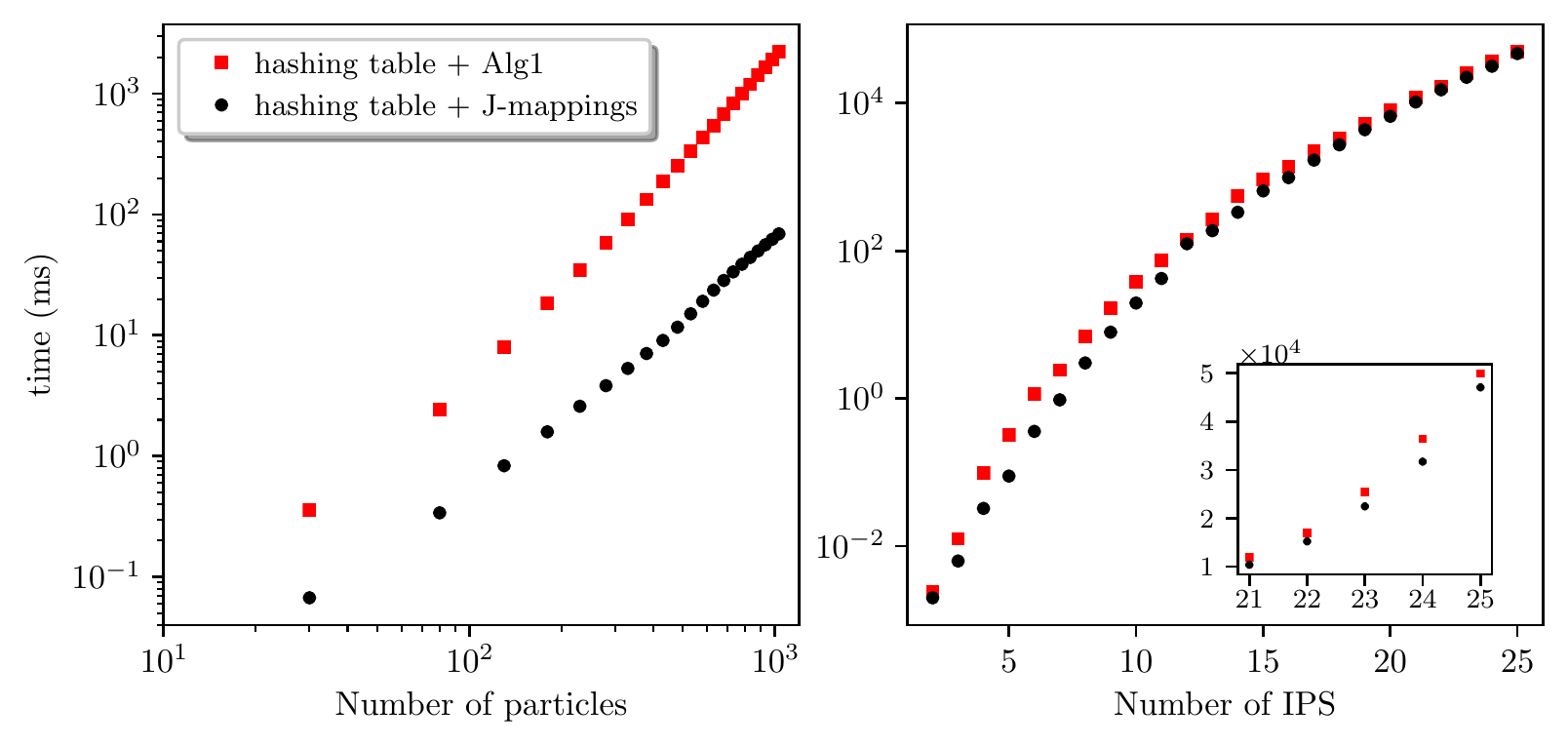}
    \caption{Time to compute all elements of $\rho^{(2)}$. The red squares correspond to an implementation that uses hashing table of configurations and calls of algorithm~\ref{alg:conf2ind} while the black circles refers to one that uses direct jump mappings between configurations related by the action of the creation/annihilation operators and dismiss completely the use of algorithms~\ref{alg:ind2conf} and~\ref{alg:conf2ind}. In the left while varying the number of particles it was taken three IPS fixed and in the right five particles was used throughout the curve with respect to IPS.}
    \label{fig:rho2_maps}
\end{figure*}

In Fig.~\ref{fig:rho2_maps}, the performance is compared between two routines that setup $\rho^{(2)}$. The first, uses the hashing table of configurations and algorithm \ref{alg:conf2ind}. The second, does not use any of the algorithms of conversion between indexes and configurations, instead, uses the  hashing table and  jump mappings (both of one and two-particle jump) explained above. The hashing table is still required to exclude forbidden transition in the rules listed in appendix \ref{appendix-matrixelements}. For large number of particles, we see a good performance gain, while for large number of IPS, the gain is slight, indicating that the use of algorithm \ref{alg:conf2ind} is not the bottleneck in this case. The C code routines, corresponding to the implementations used to generate Fig.~\ref{fig:rho2_maps}, are in the file  ``twobodyMatrix.h''. Further information about how the time elapsed is measured is provided in appendix~\ref{app:programs}.

A careful inspection in algorithm \ref{alg:conf2ind} shows us that it need to remove all particles from the configuration and thus demands the total number of particles as operations. Therefore, it is expected that the gain in performance using jump mappings is bigger for large number of particles than for large number of IPS. In other words, it is harder to empty many particles from few IPS than a few particles from many IPS.

\begin{figure*}[t]
    \centering
    \includegraphics[scale=1.0]{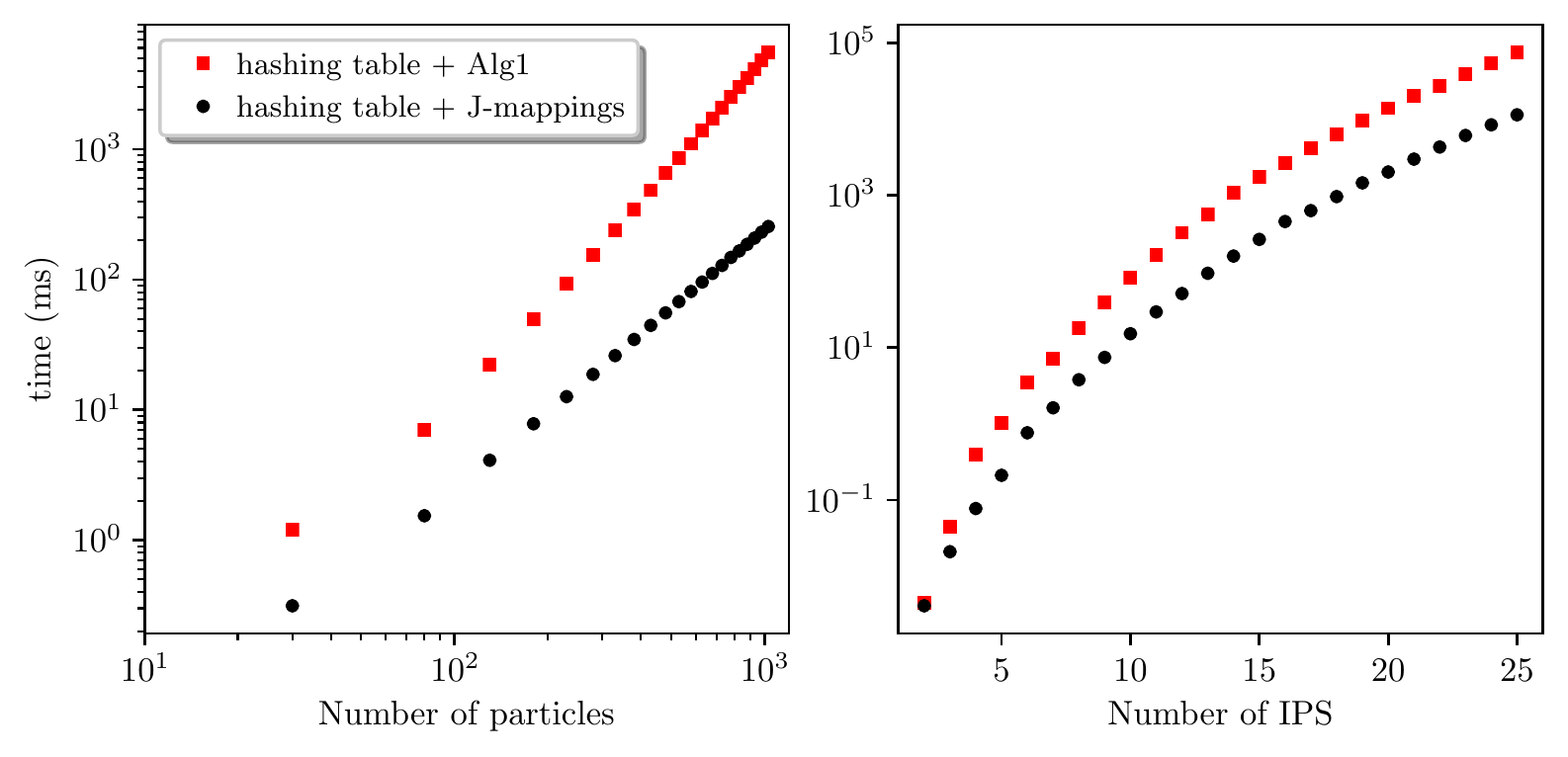}
    \caption{Time required to act with Hamiltonian operator in the configuration basis according to Eq.~\eqref{eq:Haction}. As done for $\rho^{(2)}$ in Fig.~\ref{fig:rho2_maps}, the red squares correspond to a routine that uses hashing table and algorithm \ref{alg:conf2ind} and the black circles to one that uses mappings instead of algorithm \ref{alg:conf2ind}. In the left panel we vary the number of particles for three IPS and in the right panel the number of IPS for five particles.}
    \label{fig:H_maps}
\end{figure*}

Another very important routine to check the performance gain is the time to act with the Hamiltonian over a state expressed in the multiconfigurational basis, that is, to compute $\tilde{C}$ in Eq.~\eqref{eq:Haction}. In the same way that was done for $\rho^{(2)}$, in Fig.~\ref{fig:H_maps} we compare the time demanded to compute $\tilde{C}$ using two routines, again one using the hashing table and algorithm~\ref{alg:conf2ind}, and other using the hashing table and the jump mappings. Similarly there is a clear improvement varying the number of particle (left panel), but this time there is a substantial gain also varying the number of IPS. The C code functions to act with the Hamiltonian in a vector of coefficients are in the file ``hamiltonianMatrix.h'' provided in the supplemental material.

In all comparisons between the routines that used the algorithm \ref{alg:conf2ind} with the hashing 
table and those that use mappings, when varying the number of particles, there is an evident 
constant slope behavior in the curve, at least, for large number of particles. This reveals that 
the time demanded respect to the number of particles can be written as a power law in the form

\begin{equation}
\tau_M(N) = b_M N^{a_M} ,
\label{eq:power_law}
\end{equation}

\noindent with $M$ the number of IPS fixed. The parameters can be extracted from curve fitting, 
where $a_M$ is the slope in the logarithmic scale plot. Since the study varying the number of particles in all cases presented here were carried out with $M = 3$ then we dropped the index $M$ from the parameters in the following.

\begin{table}[b]

    \centering
    {\small
    \renewcommand{\arraystretch}{1.5}
    \begin{tabular}{m{1.4cm} c c | c c}
        \toprule[2pt]
        \multicolumn{1}{c}{ } & \multicolumn{2}{c}{$\rho^{(2)}$} & 
        \multicolumn{2}{|c}{$\mathcal{H}$} \\
        \multicolumn{1}{c}{ } & $a$ & $b$ & $a$ & $b$ \\ \cline{2-5}
        hashing table & $2.835(5) $ & $6.33(9) \! \times \! 10^{-6}$ & $ 2.791(5) $ & $2.13(3) \! \times \! 10^{-5}$ \\ \hline
        jump mappings & $2.34(3) $ & $6.4(6) \! \times \! 10^{-6}$ & $2.000(2) $ & $2.40(2) \! \times \! 10^{-4}$ \\
        \bottomrule[2pt]
    \end{tabular}
    }
    
    \caption{Fitted parameters, for implementations using hashing table and jump mappings, with $M = 3$ IPS using data with $N > 100$ particles, in the left panel of Figs.~\ref{fig:rho2_maps},\ref{fig:H_maps}.}
    
    \label{tab:time_fit}
    
\end{table}

For numerical routines that compute $\rho^{(2)}$ and $\mathcal{H}$, a linear curve fitting was evaluated in the logarithmic scale plots, which resulted in a power law of the form~\eqref{eq:power_law} for time as function of the number of particles. The values are shown in Tab.~\ref{tab:time_fit}. The most important feature is that the mappings reduced the exponents for both cases of $\rho^{(2)}$ and $\mathcal{H}$, what shows that the improvement is more expressive as 
larger is the number of particles.

\begin{figure*}[t]
    \centering
    \includegraphics[scale=1.0]{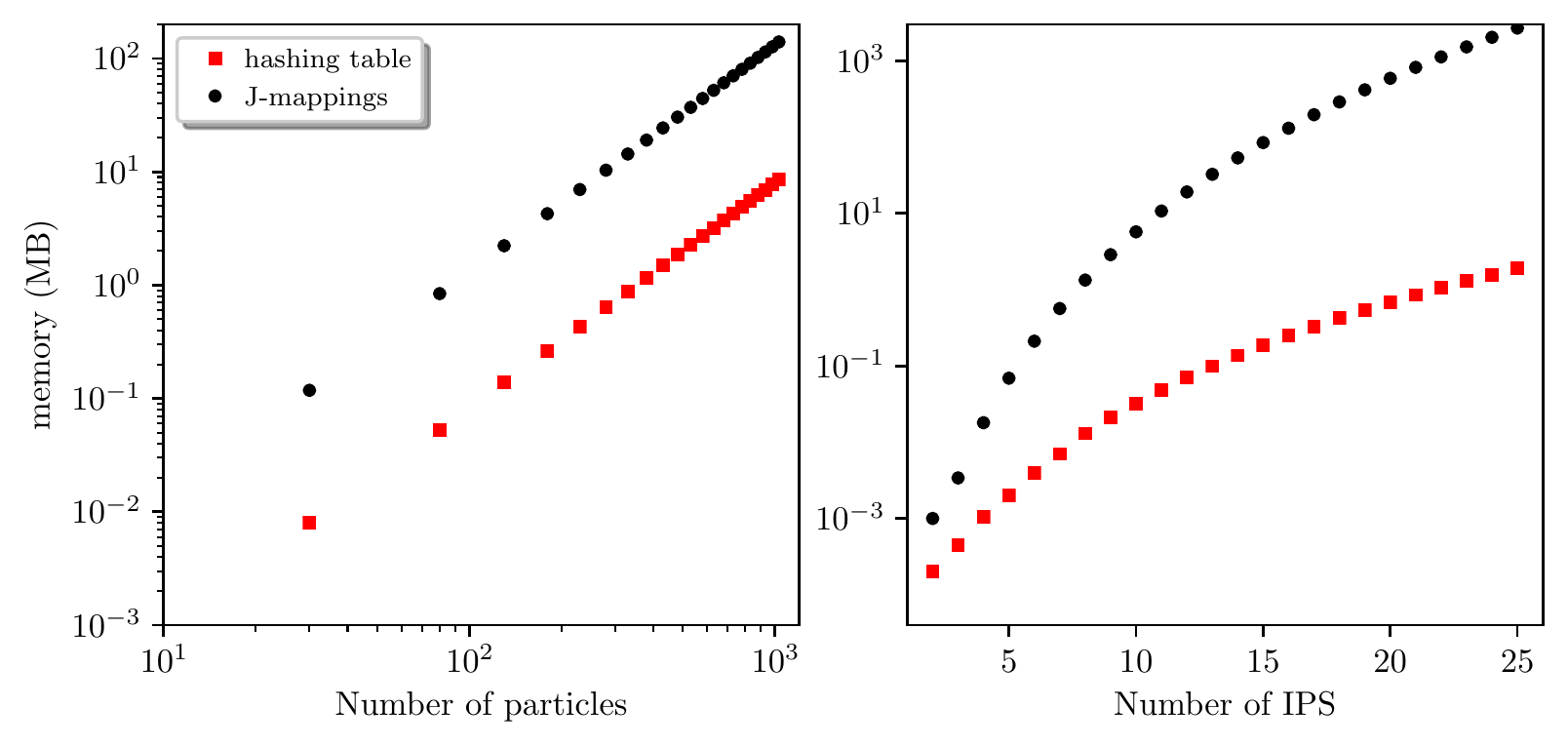}
    \caption{Memory allocation for the jump mappings structure of creation/annihilation operators and hashing table for the number of particles and IPS as used in previous figures, measured in megabytes (MB). In the left panel again we set three IPS fixed and in the right panel five particles were used. }
    \label{fig:memory}
\end{figure*}

Nevertheless, despite we have emphasized how suitable was the introduction of the jump mappings structure, we need to check the limits of application in terms of the additional memory demanded. Indeed, all the gain in time had a cost in memory, as showed in Fig.~\ref{fig:memory}. From the left panel, the case we vary the number of particles, we see that this cost is relatively cheap, some hundreds of megabytes (MB), right the case the performance gain was more expressive. The case in the right panel shows that we cannot ignore the memory consumption since it demanded up to some thousands of MB, which is not a problem for regular workstations, but indicates that a possible limitation may come up if one extrapolate $M = 25$ IPS with $N = 5$ particles.

\begin{table}[b]
    
    \centering
    \setlength{\tabcolsep}{12pt}
    \begin{tabular}{c c c}
        \toprule[2pt]
        & $a$ & $b$ \\
        hashing table & $\ 1.9953(2) \ $ & $8.288(5) \times 10^{-6}$  \\
        jump mappings & $\ 2.0004(1) \ $ & $1.3158(1) \times 10^{-4}$ \\
        \bottomrule[2pt]
    \end{tabular}
    
    \caption{Fitted parameters of power law for the memory consumption as function of $N$ for $M = 3$ fixed, for the left panel of Fig.~\ref{fig:memory}.}
    
    \label{tab:memoryfit}
    
\end{table}

Remarkably, there is a similar behavior between the memory cost in Fig.~\ref{fig:memory} 
and time execution in Fig.~\ref{fig:H_maps}, since both showed a constant slope in the logarithmic scale plot when varying the number of particles. The results of the fitting parameters according to the power law in Eq.~\eqref{eq:power_law} for memory consumption are shown in Tab.~\ref{tab:memoryfit}.

\section{Massive parallel processors application}

The use of GPUs to speed up numerical calculations is not novel. In the past decade the interest for these tools has gained attention due to their effectiveness in improvements for regular workstations, in some cases performing as fast as supercomputers. We develop specific codes using the Nvidia CUDA compiler to test the impact of massive parallelization in our routine to compute the outcome vector $\tilde{C}$ resulting from the Hamiltonian action in Eq.~\eqref{eq:Haction}, and evaluate a scalability analysis about the routine proposed here. The application of Hamiltonian in a vector expressed in the multiconfigurational basis is essential for both ED applied with Lanczos tridiagonal decomposition and the SIL integrator embedded in the MCTDHB, since this operation is required many times in both approaches.

\begin{figure*}[t]
    \centering
    \includegraphics[scale=1.0]{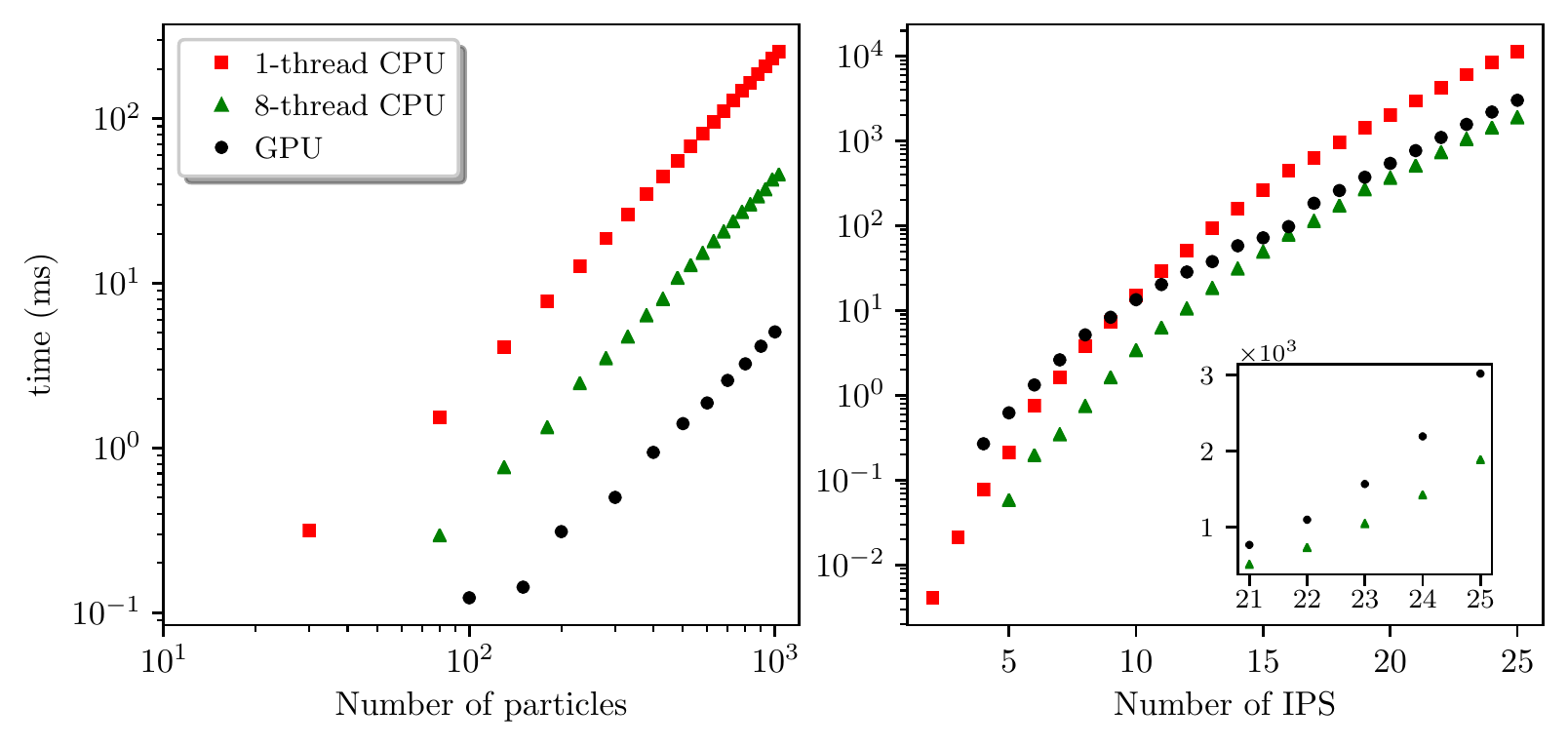}
    \caption{Performance experiment using single thread (red squares) and 8 threads (green triangles) of CPU and variable number of threads of GPU using CUDA (black circles). Just as before in the left panel was used three IPS and in the right five particles. The inset shows in normal scale the difference between CUDA and CPU parallelization for a region they are close in log scale plot.}
    \label{fig:gpu}
\end{figure*}

With the same parameters that were chosen throughout this article, in Fig.~\ref{fig:gpu}, the time elapsed computing $\tilde{\mathbf{C}} = \mathcal{H} \cdot \mathbf{C}$ is compared using one and eight cores from a CPU with the full capability of a GPU, in all cases with the mappings described in the previous section. For codes running in GPU we dynamically chose the number of blocks of threads, each one with 256 threads, to optimally exploit the GPU architecture, depending on the multiconfigurational space size $N_c(N,M)$. Additional information about the hardware used is provided in appendix~\ref{appendix-tech}.

Left panel in Fig.~\ref{fig:gpu} shows that the function is highly scalable for a large number of particles, since the presence of more threads improved dramatically the performance. Nevertheless, for the right panel we see that the scalability is much smaller for a large number of IPS. An explanation for this result lies in the amount of work each thread perform. From Eq.~\eqref{eq:Haction}, for each thread is assigned a set of indexes $\{\gamma_i\}$ to compute the $C_{\gamma_i}$ components of the vector, where the total work is balanced over all threads. However, a configuration $\vec{n}^{\gamma}$ may have many empty IPS, in which case, the specific thread responsible for this configuration will have much less work compared to another which has the particles spread over the IPS. It is worth remembering that for each new occupied IPS the number of possible jumps due to action of creation and annihilation operators scale as $M$ for single jumps and as $M^2$ for the double jumps. Therefore, the scalability in our implementation depends on a filling factor, that is, how many particles there are by IPS. 

The parallelized code routine to compute the action of the Hamiltonian is ``applyHconf\_omp'' in ``hamiltonianMatrix.h'' file, provided in the supplemental material. The parallelization for multi CPU threads is done using the OpenMP API, as the name of the function already suggests. The codes that run on GPU are separated in the ``cuda/'' folder. Specifically, to generate the data for Fig.~\ref{fig:gpu}, the function implemented in ``cuda/hamiltonianMatrix.cuh'' file is used.

\section{Physical application - Lieb-Liniger gas}

So far we reported the time demanded to compute specific operators commonly encountered in many-particles physics and did not show the results for a specific physical system. As mentioned before, the numerical routines studied here are of main importance to MCTDHB and to perform diagonalization using approximate methods like Lanczos, which require to apply the Hamiltonian in a vector expressed in the multiconfigurational basis.

In order to avoid the variational approach to the IPS which is not the main concern in this paper, we apply ED with Lanczos tridiagonal decomposition to compute approximately the ground state energy of a LL gas~\cite{PhysRev.130.1605,PhysRev.130.1616}, aiming a consistency test for our codes. The LL Hamiltonian convention used here in Schr\"odinger formalism is taken from~\cite{PhysRev.130.1605,PhysRev.130.1616,Sato_2016,SciPostPhys.3.1.003,PhysRevA.72.033613}, as
\begin{multline}
\left[ - \frac{\hbar^2}{2m} \sum_{i = 1}^{N} \frac{\partial^2}{\partial x_i^2} + 
g \sum_{i,j>i}^{N} \delta(x_j - x_i) \right] \! \psi(x_1,...,x_N)
\\ = E \psi(x_1,...,x_N) ,
\label{eq:LLHamiltonian}
\end{multline}

\noindent where the wave function is subject to periodic boundary conditions $\psi(x_1,...,x_k + L,...x_N) =
\psi(x_1,...,x_k,...x_N)$, $\forall k = 1,...,N$ and $g$ is the contact interaction strength.

The energies can be computed indirectly through the solution of a system of nonlinear equations. Here we adopt the same convention of Refs. \cite{Sato_2016,SciPostPhys.3.1.003} for these equations, which can be written as
\begin{multline}
\left\{k_j = \frac{2 \pi}{L} I_j - \frac{2}{L} \sum_{i = 1}^{N} \arctan{\left(\frac{k_j-k_i}{m g / \hbar^2}\right) } \right. ; 
\\ \left. j \in \left\{ -\frac{N-1}{2}, ..., \frac{N-1}{2} \right\} \right\} ,
\label{eq:LLsystem}
\end{multline}

\noindent where for the ground state energy we must take $I_j = j - (N+1)/2$, and the energy is related to the numbers $k_j$ by
\begin{equation}
E_0^{LL}(g) = \frac{\hbar^2}{2 m} \sum_{j = 1}^{N} k_j^2 .
\label{eq:LLE0}
\end{equation}

Despite the relation in Eq.~\eqref{eq:LLE0} looks like the kinetic energy of a ideal gas, it is worth to remind that these $k_j$ depend on the contact interaction parameter $g$. However, in the limit $g \rightarrow \infty$, it is well known that the LL gas will be described by a TG gas~\cite{doi:10.1063/1.1703687}, whose solution indicate that the energy will be given by the corresponding ideal Fermi gas, but with some care in choosing the momentum in the Slater determinant, because for a odd or even number of particles the fermions wave function must satisfy periodic or anti-periodic boundary conditions, respectively \cite{Yukalov_2005}.

\begin{figure}[b]
    \centering
    \includegraphics[scale=1]{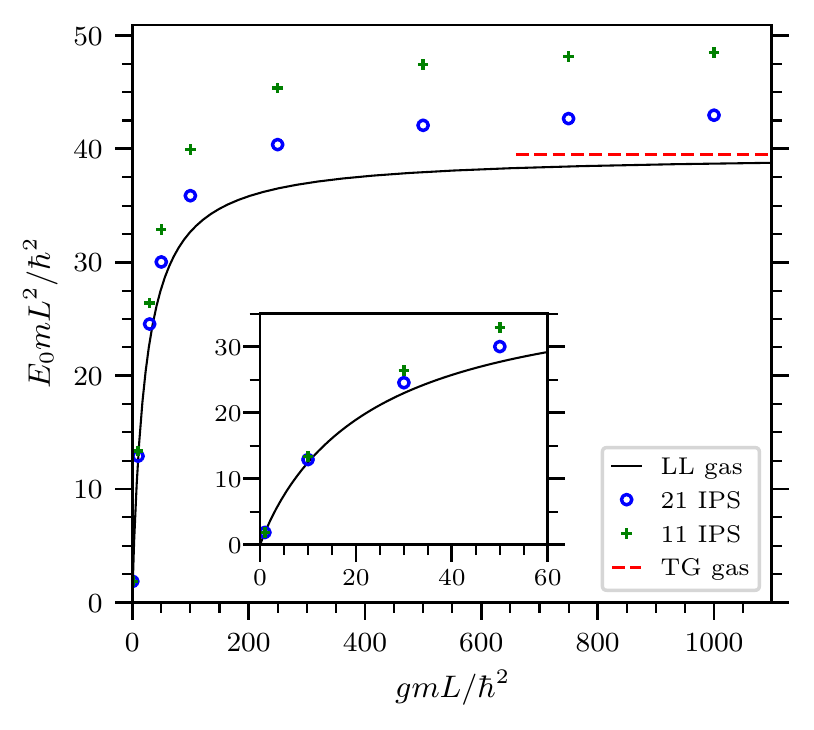}
    \caption{Ground state energy($E_0$) of the LL gas with five particles computed using numerical diagonalization (crosses and open circles) and the analytical form (full line).}
    \label{fig:LiebComparison}
\end{figure}

In Fig.~\ref{fig:LiebComparison} we compare the ground state energy obtained from numerical diagonalization for 5 particles limited to 11 and 21 IPS with the exact analytical solution computed from Eqs. (\ref{eq:LLsystem}, \ref{eq:LLE0}). The IPS in the numerical computation were chosen as periodic plane waves $\phi_n = e^{i k_n x} / \sqrt{L}$ where $k_n = 2 \pi n / L$ with  $n=-5,-4,...,4,5$ (11 IPS) and $n = -10, -9,..., 9, 10$ (21 IPS). The numerical diagonalization was computed approximately using Lanczos algorithm for tridiagonal decomposition \cite{Lanczos:1950,Loan,Demmel.ch7} together with LAPACK library~\cite{LAPACKlibrary} to diagonalize the resulting tridiagonal matrix.

The specific implementation of Lanczos iteration was done with a complete re-orthogonalization, to enforce the orthogonality of the output eigenvectors~\cite{Paige1970,SIMON1984101}, and the number of iteration was restricted to 1/8 of the dimension of the multiconfigurational space given by Eq.~\eqref{eq:nc}, since Lanczos algorithm offer good precision with small number of iterations for the smallest eigenvalue as showed in Refs. \cite{Weibe2008,Demmel.ch7}.

We can see that the deviation of the numerical solution from the exact analytical one in Fig.~\ref{fig:LiebComparison} increases with the interaction and decreases with the number of IPS. From the TG gas solution, the wave-function is not equal to the corresponding wave-function for the system of ideal fermions. Instead, it is necessary a symmetrization function, since the problem refers to bosons, as pointed out in Refs.~\cite{doi:10.1063/1.1703687,Yukalov_2005}. Therefore, despite the fermions occupy exactly 5 IPS in the ground state, this is not true for the bosons, what can be at first sight counter-intuitive. Nevertheless, our codes shows to be consistent because the approximation is better as larger is the multiconfigurational space, and besides, show quantitatively the deviation.

The C code used to solve approximately the diagonalization problem is given in file "groundStateLieb.c" provided in the supplemental material. Additionally, it requires the LAPACK library~\cite{LAPACKlibrary}.

\section{Conclusion and Outlook}

In this article, we brought different ways to implement an effective indexing of configurations 
to represent a many-particle state, which is of main concern for developing numerical 
multiconfigurational methods. We generalize the problem assuming that all particles in any 
individual particle state interact with each other, without restricting to a lattice with 
just nearest neighbor interaction or focusing in spin system. Therefore the time demanded 
exposed here is an upper bound for any bosonic system.

It was discussed carefully the performance and limitations of the different ways to build
routines for the main physical quantities, and how direct mappings of indexes can be done to track 
the action of creation/annihilation operators. The limits of applicability with the mappings 
structure is explored in terms of memory required and our results shows to be reproducible in
workstations with none special configuration.

Beyond developing the algorithms, we carried out a study of the impact from massive parallelization using GPU and compared with CPU, revealing details about the scalability of the implementation as well. For the most demanding cases, for roughly 1000 particles the best improvement was achieved by the GPU with a time reduction by a factor 50  when compared to single thread, whereas for 25 IPS the best result was with 8-threaded CPU with approximately a time reduction by a factor 6 when compared to single thread execution. Therefore, our codes showed a higher level of scalability when the number of particles is much larger than the number of IPS, since in this comparison the amount of threads provided by GPU drastically improved the performance.

Finally, we applied our routines to extract the ground state energy of the Lieb-Liniger gas and the numerical results were compared with exact analytical ones, demonstrating the correctness and the limitations of our codes. Particularly for this problem, the deviation
from the exact analytical solution is studied varying the interaction strength and number of IPS, from where we conclude they are closely related, since as the interaction strength is increased we need a larger multiconfigurational space with more IPS to better approximate the exact analytical result.

The codes presented here are very general and contains other physical systems as particular cases. One example is the Bose-Hubbard model, where the IPS are restricted to localized sites and the creation/annihilation operators appears only among neighboring sites. Thus, beyond the direct application to the Lieb-Linigar gas, one may simplify the present codes to work in different models.
Besides, no performance study about multiconfigurational methods has been worked out in such details to the best of our knowledge.

\section*{acknowledgments}

The authors thank the Brazilian agencies Fundaç\~ao de Amparo \`a Pesquisa do Estado de S\~ao Paulo (FAPESP), grant numbers 2018/02737-4 and 2016/17612-7, and Conselho Nacional de Desenvolvimento Cient\'ifico e Tecnol\'ogico (CNPq), grant number 306920/2018-2, for the project funding.

\appendix

\section{Technical Details of the Resources}
\label{appendix-tech}

The CPU used was an Intel\textsuperscript{\tiny\textregistered} 
Xeon\textsuperscript{\tiny\textregistered} CPU E5-2620 v4, clock rate 2.10GHz and 8 cores with the codes compiled with the Intel C compiler version 19. The CPU parallelization were done using the OpenMP API version 4.5. The GPU used for simulation was a NVIDIA Tesla K40c, with CUDA compiler version 10.1. We stress that in the GPU architecture the threads are divided in blocks and we by default used 256 threads per block and define the number of blocks dynamically accordingly to the size of the configurational space, trying to keep one operation per thread up to the maximum number of threads available.

\section{Matrix elements for the reduced two-body density matrices }
\label{appendix-matrixelements}

We follow Ref.~\cite{MCTDHBderivation}, and  present the reduced one-body and two-body density matrices explicitly, addapted to our notation. Starting with a $\beta$ configuration,  $\beta_a^b$ is a resulting configuration index where one particle from the $a$-th orbital is removed and then added to the $b$-th orbital. Analogously, starting from a $\beta$ configuration, $\beta_{ab}^{cd}$ is a resulting configuration index where, two particles are removed from the $a$-th and $b$-th orbitals and then added to the $c$-th and $d$-th orbital, respectively. Sums over $\beta$ index ranges from 1 to $N_c$.

\begin{align*}
\rho_{kk}&=\sum_{\beta}^{N_c}C_\beta^{*}C_{\beta} n_k, \\
\rho_{ksks}&=\sum_\beta^{N_c} C^{*}_\beta C_\beta n_k n_s, \\
\rho_{kl}&=\sum_\beta^{N_c}C_\beta^{*}C_{\beta_{k}^l}\sqrt {(n_{l}+1)n_k}, \\
\rho_{kkqq}&=\sum_\beta^{N_c} C^{*}_\beta C_{\beta_{kk}^{qq}}, \\
\rho_{kkkk}&=\sum_\beta^{N_c} C^{*}_\beta C_\beta (n_k^2-n_k), \\
\rho_{kkql}&=\sum_\beta^{N_c}  C^{*}_\beta C_{\beta_{kk}^{ql}} \sqrt{(n_k-1) n_k(n_q+1)(n_l+1)}, \\
\rho_{kkkl}&=\sum_\beta^{N_c}  C^{*}_\beta C_{\beta_{k}^{l}} (n_k-1) \sqrt{n_k(n_l+1)}, \\
\rho_{ksqq}&=\sum_\beta^{N_c}  C^{*}_\beta C_{\beta_{ks}^{qq}} \sqrt{n_k n_s(n_q+1) (n_q+2)}, \\
\rho_{ksss}&=\sum_\beta^{N_c}  C^{*}_\beta C_{\beta_{k}^{s}} n_s\sqrt{n_k(n_s+1)}, \\
\rho_{kssl}&=\sum_\beta^{N_c}  C^{*}_\beta C_{\beta_{k}^{l}} n_s\sqrt{n_k(n_l+1)}, \\
\rho_{ksql}&=\sum_\beta^{N_c}  C^{*}_\beta C_{\beta_{ks}^{ql}} \sqrt{n_k n_s (n_q+1)(n_l+1)}. \\
\end{align*}

\section{\bf Program Summary of the Supplemental Material}
\label{app:programs}

\noindent
\textbf{Program Title:} (1) demonstrateFockMap.c (2) performanceTest.c (3) performanceTest.cu (4) groundStateLieb.c \\
\textbf{Licensing provisions:} GPLv3 \\
\textbf{Programming language:} C, OpenMP C, Cuda. The C programs were tested with the GNU and Intel compilers, version 7.4 and 19.0 respectively. The programs using graphics cards were tested with the NVIDIA Cuda compiler version 10.1  \\
\textbf{Nature of problem:} (1) Find a perfect hashing function to index the possible configurations of particles in individual particle states and mappings among these configurations related by the action of creation and annihilation operators. (2) Minimize the time to assemble the one- and two-body density matrices and to act with the Hamiltonian operator. (3) Study the scalability of the routines under massive parallelization. (4) Work out a specific physical many-body problem with analytical solution to test the codes. \\
\textbf{Solution method:} (1) We employ a perfect hashing function for bosons to map configurations to integer numbers based on a combinatorial problem and create structures to track the configurations resulting from the action of one or two pairs of creation and annihilation operators in any other configuration. (2) We developed different implementations to improve performance each one using specific structures to deal with creation and annihilation operators. (3) We developed an implementation to compute the action of the Hamiltonian operator using Graphics Processor Units which provides a huge amount of threads to parallelize. (4) Compute using approximate diagonalization by Lanczos iterative method the ground state energy of the Lieb-Liniger gas. \\
\textbf{Header files:} The programs described above make use of additional files that were discussed in the text. These header files can not be compiled and contains auxiliary functions used in the programs. They have .h extension.

\textbf{Find codes online:} For the pre-print version, perhaps the codes could not be attached as supplemental material. Nevertheless, they can be consulted at~\cite{supplcodes}.

\bibliographystyle{apsrev4-1}
\bibliography{ref}

\begin{thebibliography}{50}%
\makeatletter
\providecommand \@ifxundefined [1]{%
 \@ifx{#1\undefined}
}%
\providecommand \@ifnum [1]{%
 \ifnum #1\expandafter \@firstoftwo
 \else \expandafter \@secondoftwo
 \fi
}%
\providecommand \@ifx [1]{%
 \ifx #1\expandafter \@firstoftwo
 \else \expandafter \@secondoftwo
 \fi
}%
\providecommand \natexlab [1]{#1}%
\providecommand \enquote  [1]{``#1''}%
\providecommand \bibnamefont  [1]{#1}%
\providecommand \bibfnamefont [1]{#1}%
\providecommand \citenamefont [1]{#1}%
\providecommand \href@noop [0]{\@secondoftwo}%
\providecommand \href [0]{\begingroup \@sanitize@url \@href}%
\providecommand \@href[1]{\@@startlink{#1}\@@href}%
\providecommand \@@href[1]{\endgroup#1\@@endlink}%
\providecommand \@sanitize@url [0]{\catcode `\\12\catcode `\$12\catcode
  `\&12\catcode `\#12\catcode `\^12\catcode `\_12\catcode `\%12\relax}%
\providecommand \@@startlink[1]{}%
\providecommand \@@endlink[0]{}%
\providecommand \url  [0]{\begingroup\@sanitize@url \@url }%
\providecommand \@url [1]{\endgroup\@href {#1}{\urlprefix }}%
\providecommand \urlprefix  [0]{URL }%
\providecommand \Eprint [0]{\href }%
\providecommand \doibase [0]{http://dx.doi.org/}%
\providecommand \selectlanguage [0]{\@gobble}%
\providecommand \bibinfo  [0]{\@secondoftwo}%
\providecommand \bibfield  [0]{\@secondoftwo}%
\providecommand \translation [1]{[#1]}%
\providecommand \BibitemOpen [0]{}%
\providecommand \bibitemStop [0]{}%
\providecommand \bibitemNoStop [0]{.\EOS\space}%
\providecommand \EOS [0]{\spacefactor3000\relax}%
\providecommand \BibitemShut  [1]{\csname bibitem#1\endcsname}%
\let\auto@bib@innerbib\@empty
\bibitem [{\citenamefont {Lieb}\ and\ \citenamefont
  {Liniger}(1963)}]{PhysRev.130.1605}%
  \BibitemOpen
  \bibfield  {author} {\bibinfo {author} {\bibfnamefont {E.~H.}\ \bibnamefont
  {Lieb}}\ and\ \bibinfo {author} {\bibfnamefont {W.}~\bibnamefont {Liniger}},\
  }\href {\doibase 10.1103/PhysRev.130.1605} {\bibfield  {journal} {\bibinfo
  {journal} {Phys. Rev.}\ }\textbf {\bibinfo {volume} {130}},\ \bibinfo {pages}
  {1605} (\bibinfo {year} {1963})}\BibitemShut {NoStop}%
\bibitem [{\citenamefont {Lieb}(1963)}]{PhysRev.130.1616}%
  \BibitemOpen
  \bibfield  {author} {\bibinfo {author} {\bibfnamefont {E.~H.}\ \bibnamefont
  {Lieb}},\ }\href {\doibase 10.1103/PhysRev.130.1616} {\bibfield  {journal}
  {\bibinfo  {journal} {Phys. Rev.}\ }\textbf {\bibinfo {volume} {130}},\
  \bibinfo {pages} {1616} (\bibinfo {year} {1963})}\BibitemShut {NoStop}%
\bibitem [{\citenamefont {Lang}\ \emph {et~al.}(2017)\citenamefont {Lang},
  \citenamefont {Hekking},\ and\ \citenamefont
  {Minguzzi}}]{SciPostPhys.3.1.003}%
  \BibitemOpen
  \bibfield  {author} {\bibinfo {author} {\bibfnamefont {G.}~\bibnamefont
  {Lang}}, \bibinfo {author} {\bibfnamefont {F.}~\bibnamefont {Hekking}}, \
  and\ \bibinfo {author} {\bibfnamefont {A.}~\bibnamefont {Minguzzi}},\ }\href
  {\doibase 10.21468/SciPostPhys.3.1.003} {\bibfield  {journal} {\bibinfo
  {journal} {SciPost Phys.}\ }\textbf {\bibinfo {volume} {3}},\ \bibinfo
  {pages} {003} (\bibinfo {year} {2017})}\BibitemShut {NoStop}%
\bibitem [{\citenamefont {Sakmann}\ \emph {et~al.}(2005)\citenamefont
  {Sakmann}, \citenamefont {Streltsov}, \citenamefont {Alon},\ and\
  \citenamefont {Cederbaum}}]{PhysRevA.72.033613}%
  \BibitemOpen
  \bibfield  {author} {\bibinfo {author} {\bibfnamefont {K.}~\bibnamefont
  {Sakmann}}, \bibinfo {author} {\bibfnamefont {A.~I.}\ \bibnamefont
  {Streltsov}}, \bibinfo {author} {\bibfnamefont {O.~E.}\ \bibnamefont {Alon}},
  \ and\ \bibinfo {author} {\bibfnamefont {L.~S.}\ \bibnamefont {Cederbaum}},\
  }\href {\doibase 10.1103/PhysRevA.72.033613} {\bibfield  {journal} {\bibinfo
  {journal} {Phys. Rev. A}\ }\textbf {\bibinfo {volume} {72}},\ \bibinfo
  {pages} {033613} (\bibinfo {year} {2005})}\BibitemShut {NoStop}%
\bibitem [{\citenamefont {Sato}\ \emph {et~al.}(2016)\citenamefont {Sato},
  \citenamefont {Kanamoto}, \citenamefont {Kaminishi},\ and\ \citenamefont
  {Deguchi}}]{Sato_2016}%
  \BibitemOpen
  \bibfield  {author} {\bibinfo {author} {\bibfnamefont {J.}~\bibnamefont
  {Sato}}, \bibinfo {author} {\bibfnamefont {R.}~\bibnamefont {Kanamoto}},
  \bibinfo {author} {\bibfnamefont {E.}~\bibnamefont {Kaminishi}}, \ and\
  \bibinfo {author} {\bibfnamefont {T.}~\bibnamefont {Deguchi}},\ }\href
  {\doibase 10.1088/1367-2630/18/7/075008} {\bibfield  {journal} {\bibinfo
  {journal} {New Journal of Physics}\ }\textbf {\bibinfo {volume} {18}},\
  \bibinfo {pages} {075008} (\bibinfo {year} {2016})}\BibitemShut {NoStop}%
\bibitem [{\citenamefont {Girardeau}(1960)}]{doi:10.1063/1.1703687}%
  \BibitemOpen
  \bibfield  {author} {\bibinfo {author} {\bibfnamefont {M.}~\bibnamefont
  {Girardeau}},\ }\href {\doibase 10.1063/1.1703687} {\bibfield  {journal}
  {\bibinfo  {journal} {Journal of Mathematical Physics}\ }\textbf {\bibinfo
  {volume} {1}},\ \bibinfo {pages} {516} (\bibinfo {year} {1960})},\ \Eprint
  {http://arxiv.org/abs/https://doi.org/10.1063/1.1703687}
  {https://doi.org/10.1063/1.1703687} \BibitemShut {NoStop}%
\bibitem [{\citenamefont {Yukalov}\ and\ \citenamefont
  {Girardeau}(2005)}]{Yukalov_2005}%
  \BibitemOpen
  \bibfield  {author} {\bibinfo {author} {\bibfnamefont {V.~I.}\ \bibnamefont
  {Yukalov}}\ and\ \bibinfo {author} {\bibfnamefont {M.~D.}\ \bibnamefont
  {Girardeau}},\ }\href {\doibase 10.1002/lapl.200510011} {\bibfield  {journal}
  {\bibinfo  {journal} {Laser Physics Letters}\ }\textbf {\bibinfo {volume}
  {2}},\ \bibinfo {pages} {375} (\bibinfo {year} {2005})}\BibitemShut {NoStop}%
\bibitem [{\citenamefont {Cominotti}\ \emph {et~al.}(2014)\citenamefont
  {Cominotti}, \citenamefont {Rossini}, \citenamefont {Rizzi}, \citenamefont
  {Hekking},\ and\ \citenamefont {Minguzzi}}]{AnnaMinguzzi}%
  \BibitemOpen
  \bibfield  {author} {\bibinfo {author} {\bibfnamefont {M.}~\bibnamefont
  {Cominotti}}, \bibinfo {author} {\bibfnamefont {D.}~\bibnamefont {Rossini}},
  \bibinfo {author} {\bibfnamefont {M.}~\bibnamefont {Rizzi}}, \bibinfo
  {author} {\bibfnamefont {F.}~\bibnamefont {Hekking}}, \ and\ \bibinfo
  {author} {\bibfnamefont {A.}~\bibnamefont {Minguzzi}},\ }\href {\doibase
  10.1103/PhysRevLett.113.025301} {\bibfield  {journal} {\bibinfo  {journal}
  {Phys. Rev. Lett.}\ }\textbf {\bibinfo {volume} {113}},\ \bibinfo {pages}
  {025301} (\bibinfo {year} {2014})}\BibitemShut {NoStop}%
\bibitem [{\citenamefont {Anderson}(1972)}]{Anderson1972}%
  \BibitemOpen
  \bibfield  {author} {\bibinfo {author} {\bibfnamefont {P.~W.}\ \bibnamefont
  {Anderson}},\ }\href {\doibase 10.1126/science.177.4047.393} {\bibfield
  {journal} {\bibinfo  {journal} {Science}\ }\textbf {\bibinfo {volume}
  {177}},\ \bibinfo {pages} {393} (\bibinfo {year} {1972})},\ \Eprint
  {http://arxiv.org/abs/https://science.sciencemag.org/content/177/4047/393.full.pdf}
  {https://science.sciencemag.org/content/177/4047/393.full.pdf} \BibitemShut
  {NoStop}%
\bibitem [{Note1()}]{Note1}%
  \BibitemOpen
  \bibinfo {note} {Methods employed here are also used in chemistry, from which
  came the designation for orbitals used on the study of
  molecules.}\BibitemShut {Stop}%
\bibitem [{\citenamefont {Gersch}\ and\ \citenamefont
  {Knollman}(1963)}]{AncientBoseHubbard}%
  \BibitemOpen
  \bibfield  {author} {\bibinfo {author} {\bibfnamefont {H.~A.}\ \bibnamefont
  {Gersch}}\ and\ \bibinfo {author} {\bibfnamefont {G.~C.}\ \bibnamefont
  {Knollman}},\ }\href {\doibase 10.1103/PhysRev.129.959} {\bibfield  {journal}
  {\bibinfo  {journal} {Phys. Rev.}\ }\textbf {\bibinfo {volume} {129}},\
  \bibinfo {pages} {959} (\bibinfo {year} {1963})}\BibitemShut {NoStop}%
\bibitem [{\citenamefont {Jaksch}\ and\ \citenamefont
  {Zoller}(2005)}]{JAKSCH200552}%
  \BibitemOpen
  \bibfield  {author} {\bibinfo {author} {\bibfnamefont {D.}~\bibnamefont
  {Jaksch}}\ and\ \bibinfo {author} {\bibfnamefont {P.}~\bibnamefont
  {Zoller}},\ }\href {\doibase https://doi.org/10.1016/j.aop.2004.09.010}
  {\bibfield  {journal} {\bibinfo  {journal} {Annals of Physics}\ }\textbf
  {\bibinfo {volume} {315}},\ \bibinfo {pages} {52 } (\bibinfo {year}
  {2005})},\ \bibinfo {note} {special Issue}\BibitemShut {NoStop}%
\bibitem [{\citenamefont {Jaksch}\ \emph {et~al.}(1998)\citenamefont {Jaksch},
  \citenamefont {Bruder}, \citenamefont {Cirac}, \citenamefont {Gardiner},\
  and\ \citenamefont {Zoller}}]{PhysRevLett.81.3108}%
  \BibitemOpen
  \bibfield  {author} {\bibinfo {author} {\bibfnamefont {D.}~\bibnamefont
  {Jaksch}}, \bibinfo {author} {\bibfnamefont {C.}~\bibnamefont {Bruder}},
  \bibinfo {author} {\bibfnamefont {J.~I.}\ \bibnamefont {Cirac}}, \bibinfo
  {author} {\bibfnamefont {C.~W.}\ \bibnamefont {Gardiner}}, \ and\ \bibinfo
  {author} {\bibfnamefont {P.}~\bibnamefont {Zoller}},\ }\href {\doibase
  10.1103/PhysRevLett.81.3108} {\bibfield  {journal} {\bibinfo  {journal}
  {Phys. Rev. Lett.}\ }\textbf {\bibinfo {volume} {81}},\ \bibinfo {pages}
  {3108} (\bibinfo {year} {1998})}\BibitemShut {NoStop}%
\bibitem [{\citenamefont {Fisher}\ \emph {et~al.}(1989)\citenamefont {Fisher},
  \citenamefont {Weichman}, \citenamefont {Grinstein},\ and\ \citenamefont
  {Fisher}}]{AncientSuperfluifInsulator}%
  \BibitemOpen
  \bibfield  {author} {\bibinfo {author} {\bibfnamefont {M.~P.~A.}\
  \bibnamefont {Fisher}}, \bibinfo {author} {\bibfnamefont {P.~B.}\
  \bibnamefont {Weichman}}, \bibinfo {author} {\bibfnamefont {G.}~\bibnamefont
  {Grinstein}}, \ and\ \bibinfo {author} {\bibfnamefont {D.~S.}\ \bibnamefont
  {Fisher}},\ }\href {\doibase 10.1103/PhysRevB.40.546} {\bibfield  {journal}
  {\bibinfo  {journal} {Phys. Rev. B}\ }\textbf {\bibinfo {volume} {40}},\
  \bibinfo {pages} {546} (\bibinfo {year} {1989})}\BibitemShut {NoStop}%
\bibitem [{\citenamefont {K\"uhner}\ and\ \citenamefont
  {Monien}(1998)}]{PhysRevB.58.R14741}%
  \BibitemOpen
  \bibfield  {author} {\bibinfo {author} {\bibfnamefont {T.~D.}\ \bibnamefont
  {K\"uhner}}\ and\ \bibinfo {author} {\bibfnamefont {H.}~\bibnamefont
  {Monien}},\ }\href {\doibase 10.1103/PhysRevB.58.R14741} {\bibfield
  {journal} {\bibinfo  {journal} {Phys. Rev. B}\ }\textbf {\bibinfo {volume}
  {58}},\ \bibinfo {pages} {R14741} (\bibinfo {year} {1998})}\BibitemShut
  {NoStop}%
\bibitem [{\citenamefont {Bruder}\ \emph {et~al.}(1993)\citenamefont {Bruder},
  \citenamefont {Fazio},\ and\ \citenamefont {Sch\"on}}]{PhysRevB.47.342}%
  \BibitemOpen
  \bibfield  {author} {\bibinfo {author} {\bibfnamefont {C.}~\bibnamefont
  {Bruder}}, \bibinfo {author} {\bibfnamefont {R.}~\bibnamefont {Fazio}}, \
  and\ \bibinfo {author} {\bibfnamefont {G.}~\bibnamefont {Sch\"on}},\ }\href
  {\doibase 10.1103/PhysRevB.47.342} {\bibfield  {journal} {\bibinfo  {journal}
  {Phys. Rev. B}\ }\textbf {\bibinfo {volume} {47}},\ \bibinfo {pages} {342}
  (\bibinfo {year} {1993})}\BibitemShut {NoStop}%
\bibitem [{\citenamefont {Kosloff}(1988)}]{10.1021/j100319a003}%
  \BibitemOpen
  \bibfield  {author} {\bibinfo {author} {\bibfnamefont {R.}~\bibnamefont
  {Kosloff}},\ }\href {\doibase 10.1021/j100319a003} {\bibfield  {journal}
  {\bibinfo  {journal} {The Journal of Physical Chemistry}\ }\textbf {\bibinfo
  {volume} {92}},\ \bibinfo {pages} {2087} (\bibinfo {year} {1988})},\ \Eprint
  {http://arxiv.org/abs/https://doi.org/10.1021/j100319a003}
  {https://doi.org/10.1021/j100319a003} \BibitemShut {NoStop}%
\bibitem [{\citenamefont {Kotler}\ \emph {et~al.}(1988)\citenamefont {Kotler},
  \citenamefont {Nitzan},\ and\ \citenamefont {Kosloff}}]{KOTLER1988483}%
  \BibitemOpen
  \bibfield  {author} {\bibinfo {author} {\bibfnamefont {Z.}~\bibnamefont
  {Kotler}}, \bibinfo {author} {\bibfnamefont {A.}~\bibnamefont {Nitzan}}, \
  and\ \bibinfo {author} {\bibfnamefont {R.}~\bibnamefont {Kosloff}},\ }\href
  {\doibase https://doi.org/10.1016/0009-2614(88)85247-3} {\bibfield  {journal}
  {\bibinfo  {journal} {Chemical Physics Letters}\ }\textbf {\bibinfo {volume}
  {153}},\ \bibinfo {pages} {483 } (\bibinfo {year} {1988})}\BibitemShut
  {NoStop}%
\bibitem [{\citenamefont {Meyer}\ \emph {et~al.}(1990)\citenamefont {Meyer},
  \citenamefont {Manthe},\ and\ \citenamefont {Cederbaum}}]{MEYER199073}%
  \BibitemOpen
  \bibfield  {author} {\bibinfo {author} {\bibfnamefont {H.-D.}\ \bibnamefont
  {Meyer}}, \bibinfo {author} {\bibfnamefont {U.}~\bibnamefont {Manthe}}, \
  and\ \bibinfo {author} {\bibfnamefont {L.}~\bibnamefont {Cederbaum}},\ }\href
  {\doibase https://doi.org/10.1016/0009-2614(90)87014-I} {\bibfield  {journal}
  {\bibinfo  {journal} {Chemical Physics Letters}\ }\textbf {\bibinfo {volume}
  {165}},\ \bibinfo {pages} {73 } (\bibinfo {year} {1990})}\BibitemShut
  {NoStop}%
\bibitem [{\citenamefont {Waldeck}\ \emph {et~al.}(1991)\citenamefont
  {Waldeck}, \citenamefont {Campos‐Martínez},\ and\ \citenamefont
  {Coalson}}]{10.1063/1.459854}%
  \BibitemOpen
  \bibfield  {author} {\bibinfo {author} {\bibfnamefont {J.~R.}\ \bibnamefont
  {Waldeck}}, \bibinfo {author} {\bibfnamefont {J.}~\bibnamefont
  {Campos‐Martínez}}, \ and\ \bibinfo {author} {\bibfnamefont {R.~D.}\
  \bibnamefont {Coalson}},\ }\href {\doibase 10.1063/1.459854} {\bibfield
  {journal} {\bibinfo  {journal} {The Journal of Chemical Physics}\ }\textbf
  {\bibinfo {volume} {94}},\ \bibinfo {pages} {2773} (\bibinfo {year}
  {1991})},\ \Eprint {http://arxiv.org/abs/https://doi.org/10.1063/1.459854}
  {https://doi.org/10.1063/1.459854} \BibitemShut {NoStop}%
\bibitem [{\citenamefont {Manthe}\ \emph {et~al.}(1992)\citenamefont {Manthe},
  \citenamefont {Meyer},\ and\ \citenamefont {Cederbaum}}]{10.1063/1.463007}%
  \BibitemOpen
  \bibfield  {author} {\bibinfo {author} {\bibfnamefont {U.}~\bibnamefont
  {Manthe}}, \bibinfo {author} {\bibfnamefont {H.}~\bibnamefont {Meyer}}, \
  and\ \bibinfo {author} {\bibfnamefont {L.~S.}\ \bibnamefont {Cederbaum}},\
  }\href {\doibase 10.1063/1.463007} {\bibfield  {journal} {\bibinfo  {journal}
  {The Journal of Chemical Physics}\ }\textbf {\bibinfo {volume} {97}},\
  \bibinfo {pages} {3199} (\bibinfo {year} {1992})},\ \Eprint
  {http://arxiv.org/abs/https://doi.org/10.1063/1.463007}
  {https://doi.org/10.1063/1.463007} \BibitemShut {NoStop}%
\bibitem [{\citenamefont {Alon}\ \emph {et~al.}(2008)\citenamefont {Alon},
  \citenamefont {Streltsov},\ and\ \citenamefont
  {Cederbaum}}]{MCTDHBderivation}%
  \BibitemOpen
  \bibfield  {author} {\bibinfo {author} {\bibfnamefont {O.~E.}\ \bibnamefont
  {Alon}}, \bibinfo {author} {\bibfnamefont {A.~I.}\ \bibnamefont {Streltsov}},
  \ and\ \bibinfo {author} {\bibfnamefont {L.~S.}\ \bibnamefont {Cederbaum}},\
  }\href {\doibase 10.1103/PhysRevA.77.033613} {\bibfield  {journal} {\bibinfo
  {journal} {Phys. Rev. A}\ }\textbf {\bibinfo {volume} {77}},\ \bibinfo
  {pages} {033613} (\bibinfo {year} {2008})}\BibitemShut {NoStop}%
\bibitem [{\citenamefont {Lode}\ \emph {et~al.}(2012)\citenamefont {Lode},
  \citenamefont {Sakmann}, \citenamefont {Alon}, \citenamefont {Cederbaum},\
  and\ \citenamefont {Streltsov}}]{HarmonicInteractModel}%
  \BibitemOpen
  \bibfield  {author} {\bibinfo {author} {\bibfnamefont {A.~U.~J.}\
  \bibnamefont {Lode}}, \bibinfo {author} {\bibfnamefont {K.}~\bibnamefont
  {Sakmann}}, \bibinfo {author} {\bibfnamefont {O.~E.}\ \bibnamefont {Alon}},
  \bibinfo {author} {\bibfnamefont {L.~S.}\ \bibnamefont {Cederbaum}}, \ and\
  \bibinfo {author} {\bibfnamefont {A.~I.}\ \bibnamefont {Streltsov}},\ }\href
  {\doibase 10.1103/PhysRevA.86.063606} {\bibfield  {journal} {\bibinfo
  {journal} {Phys. Rev. A}\ }\textbf {\bibinfo {volume} {86}},\ \bibinfo
  {pages} {063606} (\bibinfo {year} {2012})}\BibitemShut {NoStop}%
\bibitem [{\citenamefont {Klaiman}\ and\ \citenamefont
  {Cederbaum}(2016)}]{PhysRevA.94.063648}%
  \BibitemOpen
  \bibfield  {author} {\bibinfo {author} {\bibfnamefont {S.}~\bibnamefont
  {Klaiman}}\ and\ \bibinfo {author} {\bibfnamefont {L.~S.}\ \bibnamefont
  {Cederbaum}},\ }\href {\doibase 10.1103/PhysRevA.94.063648} {\bibfield
  {journal} {\bibinfo  {journal} {Phys. Rev. A}\ }\textbf {\bibinfo {volume}
  {94}},\ \bibinfo {pages} {063648} (\bibinfo {year} {2016})}\BibitemShut
  {NoStop}%
\bibitem [{\citenamefont {Lode}\ \emph {et~al.}(2015)\citenamefont {Lode},
  \citenamefont {Chakrabarti},\ and\ \citenamefont
  {Kota}}]{PhysRevA.92.033622}%
  \BibitemOpen
  \bibfield  {author} {\bibinfo {author} {\bibfnamefont {A.~U.~J.}\
  \bibnamefont {Lode}}, \bibinfo {author} {\bibfnamefont {B.}~\bibnamefont
  {Chakrabarti}}, \ and\ \bibinfo {author} {\bibfnamefont {V.~K.~B.}\
  \bibnamefont {Kota}},\ }\href {\doibase 10.1103/PhysRevA.92.033622}
  {\bibfield  {journal} {\bibinfo  {journal} {Phys. Rev. A}\ }\textbf {\bibinfo
  {volume} {92}},\ \bibinfo {pages} {033622} (\bibinfo {year}
  {2015})}\BibitemShut {NoStop}%
\bibitem [{\citenamefont {Roy}\ \emph {et~al.}(2018)\citenamefont {Roy},
  \citenamefont {Gammal}, \citenamefont {Tsatsos}, \citenamefont {Chatterjee},
  \citenamefont {Chakrabarti},\ and\ \citenamefont
  {Lode}}]{PhysRevA.97.043625}%
  \BibitemOpen
  \bibfield  {author} {\bibinfo {author} {\bibfnamefont {R.}~\bibnamefont
  {Roy}}, \bibinfo {author} {\bibfnamefont {A.}~\bibnamefont {Gammal}},
  \bibinfo {author} {\bibfnamefont {M.~C.}\ \bibnamefont {Tsatsos}}, \bibinfo
  {author} {\bibfnamefont {B.}~\bibnamefont {Chatterjee}}, \bibinfo {author}
  {\bibfnamefont {B.}~\bibnamefont {Chakrabarti}}, \ and\ \bibinfo {author}
  {\bibfnamefont {A.~U.~J.}\ \bibnamefont {Lode}},\ }\href {\doibase
  10.1103/PhysRevA.97.043625} {\bibfield  {journal} {\bibinfo  {journal} {Phys.
  Rev. A}\ }\textbf {\bibinfo {volume} {97}},\ \bibinfo {pages} {043625}
  (\bibinfo {year} {2018})}\BibitemShut {NoStop}%
\bibitem [{\citenamefont {Lode}\ and\ \citenamefont
  {Bruder}(2017)}]{PhysRevLett.118.013603}%
  \BibitemOpen
  \bibfield  {author} {\bibinfo {author} {\bibfnamefont {A.~U.~J.}\
  \bibnamefont {Lode}}\ and\ \bibinfo {author} {\bibfnamefont {C.}~\bibnamefont
  {Bruder}},\ }\href {\doibase 10.1103/PhysRevLett.118.013603} {\bibfield
  {journal} {\bibinfo  {journal} {Phys. Rev. Lett.}\ }\textbf {\bibinfo
  {volume} {118}},\ \bibinfo {pages} {013603} (\bibinfo {year}
  {2017})}\BibitemShut {NoStop}%
\bibitem [{\citenamefont {Nguyen}\ \emph {et~al.}(2019)\citenamefont {Nguyen},
  \citenamefont {Tsatsos}, \citenamefont {Luo}, \citenamefont {Lode},
  \citenamefont {Telles}, \citenamefont {Bagnato},\ and\ \citenamefont
  {Hulet}}]{PhysRevX.9.011052}%
  \BibitemOpen
  \bibfield  {author} {\bibinfo {author} {\bibfnamefont {J.~H.~V.}\
  \bibnamefont {Nguyen}}, \bibinfo {author} {\bibfnamefont {M.~C.}\
  \bibnamefont {Tsatsos}}, \bibinfo {author} {\bibfnamefont {D.}~\bibnamefont
  {Luo}}, \bibinfo {author} {\bibfnamefont {A.~U.~J.}\ \bibnamefont {Lode}},
  \bibinfo {author} {\bibfnamefont {G.~D.}\ \bibnamefont {Telles}}, \bibinfo
  {author} {\bibfnamefont {V.~S.}\ \bibnamefont {Bagnato}}, \ and\ \bibinfo
  {author} {\bibfnamefont {R.~G.}\ \bibnamefont {Hulet}},\ }\href {\doibase
  10.1103/PhysRevX.9.011052} {\bibfield  {journal} {\bibinfo  {journal} {Phys.
  Rev. X}\ }\textbf {\bibinfo {volume} {9}},\ \bibinfo {pages} {011052}
  (\bibinfo {year} {2019})}\BibitemShut {NoStop}%
\bibitem [{\citenamefont {Klaiman}\ \emph {et~al.}(2014)\citenamefont
  {Klaiman}, \citenamefont {Lode}, \citenamefont {Streltsov}, \citenamefont
  {Cederbaum},\ and\ \citenamefont {Alon}}]{PhysRevA.90.043620}%
  \BibitemOpen
  \bibfield  {author} {\bibinfo {author} {\bibfnamefont {S.}~\bibnamefont
  {Klaiman}}, \bibinfo {author} {\bibfnamefont {A.~U.~J.}\ \bibnamefont
  {Lode}}, \bibinfo {author} {\bibfnamefont {A.~I.}\ \bibnamefont {Streltsov}},
  \bibinfo {author} {\bibfnamefont {L.~S.}\ \bibnamefont {Cederbaum}}, \ and\
  \bibinfo {author} {\bibfnamefont {O.~E.}\ \bibnamefont {Alon}},\ }\href
  {\doibase 10.1103/PhysRevA.90.043620} {\bibfield  {journal} {\bibinfo
  {journal} {Phys. Rev. A}\ }\textbf {\bibinfo {volume} {90}},\ \bibinfo
  {pages} {043620} (\bibinfo {year} {2014})}\BibitemShut {NoStop}%
\bibitem [{\citenamefont {B\ifmmode~\check{r}\else \v{r}\fi{}ezinov\'a}\ \emph
  {et~al.}(2012)\citenamefont {B\ifmmode~\check{r}\else \v{r}\fi{}ezinov\'a},
  \citenamefont {Lode}, \citenamefont {Streltsov}, \citenamefont {Alon},
  \citenamefont {Cederbaum},\ and\ \citenamefont
  {Burgd\"orfer}}]{PhysRevA.86.013630}%
  \BibitemOpen
  \bibfield  {author} {\bibinfo {author} {\bibfnamefont {I.}~\bibnamefont
  {B\ifmmode~\check{r}\else \v{r}\fi{}ezinov\'a}}, \bibinfo {author}
  {\bibfnamefont {A.~U.~J.}\ \bibnamefont {Lode}}, \bibinfo {author}
  {\bibfnamefont {A.~I.}\ \bibnamefont {Streltsov}}, \bibinfo {author}
  {\bibfnamefont {O.~E.}\ \bibnamefont {Alon}}, \bibinfo {author}
  {\bibfnamefont {L.~S.}\ \bibnamefont {Cederbaum}}, \ and\ \bibinfo {author}
  {\bibfnamefont {J.}~\bibnamefont {Burgd\"orfer}},\ }\href {\doibase
  10.1103/PhysRevA.86.013630} {\bibfield  {journal} {\bibinfo  {journal} {Phys.
  Rev. A}\ }\textbf {\bibinfo {volume} {86}},\ \bibinfo {pages} {013630}
  (\bibinfo {year} {2012})}\BibitemShut {NoStop}%
\bibitem [{\citenamefont {Fischer}\ \emph {et~al.}(2015)\citenamefont
  {Fischer}, \citenamefont {Lode},\ and\ \citenamefont
  {Chatterjee}}]{PhysRevA.91.063621}%
  \BibitemOpen
  \bibfield  {author} {\bibinfo {author} {\bibfnamefont {U.~R.}\ \bibnamefont
  {Fischer}}, \bibinfo {author} {\bibfnamefont {A.~U.~J.}\ \bibnamefont
  {Lode}}, \ and\ \bibinfo {author} {\bibfnamefont {B.}~\bibnamefont
  {Chatterjee}},\ }\href {\doibase 10.1103/PhysRevA.91.063621} {\bibfield
  {journal} {\bibinfo  {journal} {Phys. Rev. A}\ }\textbf {\bibinfo {volume}
  {91}},\ \bibinfo {pages} {063621} (\bibinfo {year} {2015})}\BibitemShut
  {NoStop}%
\bibitem [{\citenamefont {Andriati}\ and\ \citenamefont
  {Gammal}(2019)}]{PhysRevA.100.063625}%
  \BibitemOpen
  \bibfield  {author} {\bibinfo {author} {\bibfnamefont {A.~V.}\ \bibnamefont
  {Andriati}}\ and\ \bibinfo {author} {\bibfnamefont {A.}~\bibnamefont
  {Gammal}},\ }\href {\doibase 10.1103/PhysRevA.100.063625} {\bibfield
  {journal} {\bibinfo  {journal} {Phys. Rev. A}\ }\textbf {\bibinfo {volume}
  {100}},\ \bibinfo {pages} {063625} (\bibinfo {year} {2019})}\BibitemShut
  {NoStop}%
\bibitem [{\citenamefont {Wei{\ss}e}\ and\ \citenamefont
  {Fehske}(2008)}]{Weibe2008}%
  \BibitemOpen
  \bibfield  {author} {\bibinfo {author} {\bibfnamefont {A.}~\bibnamefont
  {Wei{\ss}e}}\ and\ \bibinfo {author} {\bibfnamefont {H.}~\bibnamefont
  {Fehske}},\ }in\ \href {\doibase 10.1007/978-3-540-74686-7_18} {\emph
  {\bibinfo {booktitle} {Computational Many-Particle Physics}}},\ \bibinfo
  {editor} {edited by\ \bibinfo {editor} {\bibfnamefont {H.}~\bibnamefont
  {Fehske}}, \bibinfo {editor} {\bibfnamefont {R.}~\bibnamefont {Schneider}}, \
  and\ \bibinfo {editor} {\bibfnamefont {A.}~\bibnamefont {Wei{\ss}e}}}\
  (\bibinfo  {publisher} {Springer Berlin Heidelberg},\ \bibinfo {address}
  {Berlin, Heidelberg},\ \bibinfo {year} {2008})\ pp.\ \bibinfo {pages}
  {529--544}\BibitemShut {NoStop}%
\bibitem [{\citenamefont {Zhang}\ and\ \citenamefont
  {Dong}(2010)}]{Zhang_2010}%
  \BibitemOpen
  \bibfield  {author} {\bibinfo {author} {\bibfnamefont {J.~M.}\ \bibnamefont
  {Zhang}}\ and\ \bibinfo {author} {\bibfnamefont {R.~X.}\ \bibnamefont
  {Dong}},\ }\href {\doibase 10.1088/0143-0807/31/3/016} {\bibfield  {journal}
  {\bibinfo  {journal} {European Journal of Physics}\ }\textbf {\bibinfo
  {volume} {31}},\ \bibinfo {pages} {591} (\bibinfo {year} {2010})}\BibitemShut
  {NoStop}%
\bibitem [{\citenamefont {Lin}(1990)}]{PhysRevB.42.6561}%
  \BibitemOpen
  \bibfield  {author} {\bibinfo {author} {\bibfnamefont {H.~Q.}\ \bibnamefont
  {Lin}},\ }\href {\doibase 10.1103/PhysRevB.42.6561} {\bibfield  {journal}
  {\bibinfo  {journal} {Phys. Rev. B}\ }\textbf {\bibinfo {volume} {42}},\
  \bibinfo {pages} {6561} (\bibinfo {year} {1990})}\BibitemShut {NoStop}%
\bibitem [{\citenamefont {Ravent{\'{o}}s}\ \emph {et~al.}(2017)\citenamefont
  {Ravent{\'{o}}s}, \citenamefont {Gra{\ss}}, \citenamefont {Lewenstein},\ and\
  \citenamefont {Juli{\'{a}}-D{\'{\i}}az}}]{Ravent_s_2017}%
  \BibitemOpen
  \bibfield  {author} {\bibinfo {author} {\bibfnamefont {D.}~\bibnamefont
  {Ravent{\'{o}}s}}, \bibinfo {author} {\bibfnamefont {T.}~\bibnamefont
  {Gra{\ss}}}, \bibinfo {author} {\bibfnamefont {M.}~\bibnamefont
  {Lewenstein}}, \ and\ \bibinfo {author} {\bibfnamefont {B.}~\bibnamefont
  {Juli{\'{a}}-D{\'{\i}}az}},\ }\href {\doibase 10.1088/1361-6455/aa68b1}
  {\bibfield  {journal} {\bibinfo  {journal} {Journal of Physics B: Atomic,
  Molecular and Optical Physics}\ }\textbf {\bibinfo {volume} {50}},\ \bibinfo
  {pages} {113001} (\bibinfo {year} {2017})}\BibitemShut {NoStop}%
\bibitem [{\citenamefont {Sandvik}(2010)}]{Sandvik}%
  \BibitemOpen
  \bibfield  {author} {\bibinfo {author} {\bibfnamefont {A.~W.}\ \bibnamefont
  {Sandvik}},\ }\href {\doibase 10.1063/1.3518900} {\bibfield  {journal}
  {\bibinfo  {journal} {AIP Conference Proceedings}\ }\textbf {\bibinfo
  {volume} {1297}},\ \bibinfo {pages} {135} (\bibinfo {year} {2010})},\ \Eprint
  {http://arxiv.org/abs/https://aip.scitation.org/doi/pdf/10.1063/1.3518900}
  {https://aip.scitation.org/doi/pdf/10.1063/1.3518900} \BibitemShut {NoStop}%
\bibitem [{\citenamefont {Liang}(1995)}]{LIANG199511}%
  \BibitemOpen
  \bibfield  {author} {\bibinfo {author} {\bibfnamefont {S.}~\bibnamefont
  {Liang}},\ }\href {\doibase https://doi.org/10.1016/0010-4655(95)00108-R}
  {\bibfield  {journal} {\bibinfo  {journal} {Comput. Phys. Commun.}\ }\textbf
  {\bibinfo {volume} {92}},\ \bibinfo {pages} {11 } (\bibinfo {year}
  {1995})}\BibitemShut {NoStop}%
\bibitem [{\citenamefont {Jia}\ \emph {et~al.}(2018)\citenamefont {Jia},
  \citenamefont {Wang}, \citenamefont {Mendl}, \citenamefont {Moritz},\ and\
  \citenamefont {Devereaux}}]{JIA201881}%
  \BibitemOpen
  \bibfield  {author} {\bibinfo {author} {\bibfnamefont {C.}~\bibnamefont
  {Jia}}, \bibinfo {author} {\bibfnamefont {Y.}~\bibnamefont {Wang}}, \bibinfo
  {author} {\bibfnamefont {C.}~\bibnamefont {Mendl}}, \bibinfo {author}
  {\bibfnamefont {B.}~\bibnamefont {Moritz}}, \ and\ \bibinfo {author}
  {\bibfnamefont {T.}~\bibnamefont {Devereaux}},\ }\href {\doibase
  https://doi.org/10.1016/j.cpc.2017.11.011} {\bibfield  {journal} {\bibinfo
  {journal} {Comput. Phys. Commun.}\ }\textbf {\bibinfo {volume} {224}},\
  \bibinfo {pages} {81 } (\bibinfo {year} {2018})}\BibitemShut {NoStop}%
\bibitem [{\citenamefont {Lanczos}(1950)}]{Lanczos:1950}%
  \BibitemOpen
  \bibfield  {author} {\bibinfo {author} {\bibfnamefont {C.}~\bibnamefont
  {Lanczos}},\ }\href {\doibase 10.6028/jres.045.026} {\bibfield  {journal}
  {\bibinfo  {journal} {J. Res. Natl. Bur. Stand. B}\ }\textbf {\bibinfo
  {volume} {45}},\ \bibinfo {pages} {255} (\bibinfo {year} {1950})}\BibitemShut
  {NoStop}%
\bibitem [{\citenamefont {Loan}\ and\ \citenamefont {Golub}(1996)}]{Loan}%
  \BibitemOpen
  \bibfield  {author} {\bibinfo {author} {\bibfnamefont {C.~F.~V.}\
  \bibnamefont {Loan}}\ and\ \bibinfo {author} {\bibfnamefont {G.~H.}\
  \bibnamefont {Golub}},\ }\href@noop {} {\emph {\bibinfo {title} {Matrix
  Computations}}}\ (\bibinfo  {publisher} {The Johns Hopkins University
  Press},\ \bibinfo {address} {London},\ \bibinfo {year} {1996})\BibitemShut
  {NoStop}%
\bibitem [{\citenamefont {Demmel}()}]{Demmel.ch7}%
  \BibitemOpen
  \bibfield  {author} {\bibinfo {author} {\bibfnamefont {J.~W.}\ \bibnamefont
  {Demmel}},\ }\enquote {\bibinfo {title} {7. iterative methods for eigenvalue
  problems},}\ in\ \href {\doibase 10.1137/1.9781611971446.ch7} {\emph
  {\bibinfo {booktitle} {Applied Numerical Linear Algebra}}},\ Chap.~\bibinfo
  {chapter} {7}, pp.\ \bibinfo {pages} {361--387},\ \Eprint
  {http://arxiv.org/abs/https://epubs.siam.org/doi/pdf/10.1137/1.9781611971446.ch7}
  {https://epubs.siam.org/doi/pdf/10.1137/1.9781611971446.ch7} \BibitemShut
  {NoStop}%
\bibitem [{\citenamefont {Beck}\ and\ \citenamefont
  {Meyer}(1997)}]{MeyerSpringer1997}%
  \BibitemOpen
  \bibfield  {author} {\bibinfo {author} {\bibfnamefont {M.}~\bibnamefont
  {Beck}}\ and\ \bibinfo {author} {\bibfnamefont {H.-D.}\ \bibnamefont
  {Meyer}},\ }\href {\doibase 10.1007/s004600050342} {\bibfield  {journal}
  {\bibinfo  {journal} {Zeitschrift für Physik D Atoms, Molecules and
  Clusters}\ }\textbf {\bibinfo {volume} {42}},\ \bibinfo {pages} {113}
  (\bibinfo {year} {1997})}\BibitemShut {NoStop}%
\bibitem [{\citenamefont {Park}\ and\ \citenamefont {Light}(1986)}]{ParkLight}%
  \BibitemOpen
  \bibfield  {author} {\bibinfo {author} {\bibfnamefont {T.~J.}\ \bibnamefont
  {Park}}\ and\ \bibinfo {author} {\bibfnamefont {J.~C.}\ \bibnamefont
  {Light}},\ }\href {\doibase 10.1063/1.451548} {\bibfield  {journal} {\bibinfo
   {journal} {The Journal of Chemical Physics}\ }\textbf {\bibinfo {volume}
  {85}},\ \bibinfo {pages} {5870} (\bibinfo {year} {1986})}\BibitemShut
  {NoStop}%
\bibitem [{Note2()}]{Note2}%
  \BibitemOpen
  \bibinfo {note} {Actually, this number can be halved if one uses hermiticity,
  and reduced even more using the commutation relations in the case of $\rho
  ^{(2)}$. Moreover, when the indexes of the creation and annihilation
  operators are the same no calls of the conversion algorithms are need at all,
  since there is no rearrangement of particles.}\BibitemShut {Stop}%
\bibitem [{Note3()}]{Note3}%
  \BibitemOpen
  \bibinfo {note} {Jump here means the simultaneous destruction and creation of
  particle in different states.}\BibitemShut {Stop}%
\bibitem [{LAP()}]{LAPACKlibrary}%
  \BibitemOpen
  \href@noop {} {\enquote {\bibinfo {title} {Linear algebra package},}\
  }\bibinfo {howpublished} {\url{http://www.netlib.org/lapack/}},\ \bibinfo
  {note} {accessed: 2019-02-12}\BibitemShut {NoStop}%
\bibitem [{\citenamefont {Paige}(1970)}]{Paige1970}%
  \BibitemOpen
  \bibfield  {author} {\bibinfo {author} {\bibfnamefont {C.~C.}\ \bibnamefont
  {Paige}},\ }\href {\doibase 10.1007/BF01936866} {\bibfield  {journal}
  {\bibinfo  {journal} {BIT Numerical Mathematics}\ }\textbf {\bibinfo {volume}
  {10}},\ \bibinfo {pages} {183} (\bibinfo {year} {1970})}\BibitemShut
  {NoStop}%
\bibitem [{\citenamefont {Simon}(1984)}]{SIMON1984101}%
  \BibitemOpen
  \bibfield  {author} {\bibinfo {author} {\bibfnamefont {H.~D.}\ \bibnamefont
  {Simon}},\ }\href {\doibase https://doi.org/10.1016/0024-3795(84)90025-9}
  {\bibfield  {journal} {\bibinfo  {journal} {Linear Algebra and its
  Applications}\ }\textbf {\bibinfo {volume} {61}},\ \bibinfo {pages} {101 }
  (\bibinfo {year} {1984})}\BibitemShut {NoStop}%
\bibitem [{\citenamefont {Andriati}()}]{supplcodes}%
  \BibitemOpen
  \bibfield  {author} {\bibinfo {author} {\bibfnamefont {A.~V.}\ \bibnamefont
  {Andriati}},\ }\href@noop {} {\enquote {\bibinfo {title}
  {Multiconfigurational implementation},}\ }\bibinfo {howpublished}
  {\url{http://github.com/andriati-alex/mcpi}},\ \bibinfo {note} {accessed:
  2020-02-12}\BibitemShut {NoStop}%
\end{thebibliography}%

\end{document}